\pdfoutput=1 
\documentclass[multphys,vecphys]{svmult}

\usepackage{makeidx}     
\usepackage{graphicx}    
\usepackage{multicol}    

\makeindex             
\begin{document}
%
\newcommand{\uDelta}{\mbox{\greeksym{D}}}
\newcommand{\uPi}{\mbox{\greeksym{P}}}
\newcommand{\mug}{\mbox{\greeksym{m}}}
\newcommand{\mugp}{\mbox{\greeksym{m}}}
\newcommand{\Omg}{\mbox{\greeksym{W}}}
\setcounter{secnumdepth}{3} 
\catcode`@=11
\def\chapter{\thispagestyle{empty}
   \global\@topnum\z@\@afterindentfalse
   \secdef\@chapter\@schapter}
\catcode`@=12
\def\fract#1#2{{\textstyle\frac{#1}{#2}}}
\def\fracd#1#2{{\displaystyle\frac{#1}{#2}}}
\def\beq{\begin{eqnarray*}}
\def\eeq{\end{eqnarray*}}
\def\lz{\vspace{\baselineskip}}
\def\hlz{\vspace{0.5\baselineskip}}
\def\hhlz{\vspace{0.25\baselineskip}}
\def\lzn{\vspace{\baselineskip}\noindent}
\def\hlzn{\vspace{0.5\baselineskip}\noindent}
\def\hhlzn{\vspace{0.25\baselineskip}\noindent}
\def\rlz{\vspace{-\baselineskip}}
\def\rhlz{\vspace{-0.5\baselineskip}}
\def\rhhlz{\vspace{-0.25\baselineskip}}
\def\rlzn{\vspace{-\baselineskip}\noindent}
\def\rhlzn{\vspace{-0.5\baselineskip}\noindent}
\def\rhhlzn{\vspace{-0.25\baselineskip}\noindent}
\def\tbee{\hspace*{22.25pt}} 
\def\tbez{\hspace*{29pt}}    
\def\tbze{\hspace*{29pt}}    
\def\tbec{\hspace*{17.60pt}} 
\def\tce{\hspace*{34.389pt}} 
\def\tcz{\hspace*{40.139pt}} 
\def\endd{\end{document}}
%
\overfullrule=0pt
\def\abfa{\newdimen\figgap\figgap=18pt}
\def\abfn{\newdimen\figgap\figgap=300pt}
\def\intl{\int\limits}
\newcommand{\solm}{M$_\odot$}
\newcommand{\MOLH}{H$_2$}
\newcommand{\TKIN}{T_{\rm k}}
\newcommand{\kms}{km\,s$^{-1}$}
\newcommand{\kmsMpc}{km\,s$^{-1}$\,Mpc$^{-1}$}
\newcommand{\WAT}{H$_{2}$O}
\newcommand{\FORM}{H$_{2}$CO}
\newcommand{\AMM}{NH$_3$}
\newcommand{\percc}{cm$^{-3}$}
\newcommand{\pers}{s$^{-1}$}
\newcommand{\HCOP}{HCO$^+$}
\newcommand{\HTHPL}{H$_3^+$}
\newcommand{\etal}{et al.}
\newcommand{\TEX}{T_{\rm ex}}
\newcommand{\TMB}{T_{\rm MB}}
\newcommand{\Tsys}{T_{\rm sys}}
\newcommand{\CEIO}{C$^{18}$O}
\newcommand{\percmsq}{cm$^{-2}$}
\newcommand{\METH}{CH$_3$OH}
\newcommand{\CYAN}{HC$_3$N}
\newcommand{\dvel}{$\Delta {\rm V}_{1/2}$}
\newcommand{\vlsr}{V$_{\rm lsr}$}
\newcommand{\TROT}{T_{\rm rot}}
\newcommand{\CHTHCCH}{CH$_3$CCH}
\newcommand{\CHTHCN}{CH$_3$CN}
\newcommand{\solmass}{M$_\odot$}
\newcommand{\DGC}{D$_{\rm GC}$}
\newcommand{\THCO}{$^{13}$CO}
\newcommand{\HI}{H\,\mbox{\sc i}}
\newcommand{\HII}{H\,\mbox{\sc ii}}
\newcommand{\hii}{H\,\mbox{\sc ii}}
\newcommand{\tb}{T_{\rm B}}
\newcommand{\ta}{T_{\rm A}}
\newcommand{\tsp}{T_{\rm s}}
\newcommand{\Tb}{T_{\rm B}}
\newcommand{\Ta}{T_{\rm A}}
\newcommand{\Tsp}{T_{\rm s}}
\newcommand{\te}{T_{\rm e}}
\newcommand{\expo}{{\rm \,e}}
\newcommand{\diff}{{\rm \, d}}
\newcommand{\im}{{\rm \,i\,}}
\renewcommand{\Re}{{\rm Re}}
\renewcommand{\Im}{{\rm Im}}
\newcommand{\pl}{\parallel}
\newcommand{\sha}{\raisebox{-0.2ex}{\kern1pt\rm III\,}}
\newcommand{\ddel}{\raisebox{-0.2ex}{\rm II\,}}
\newcommand{\la}{\lesssim}
\newcommand{\ga}{\gtrapprox}
\newcommand{\gequal}{\mathrel{\stackrel{>}{=}}}
\newcommand{\lequal}{\mathrel{\stackrel{<}{=}}}
\newcommand{\longpage}{\enlargethispage{\baselineskip}}
\newcommand{\shortpage}{\enlargethispage{-\baselineskip}}
\def\endd{\end{document}}
\def\enda{\end{document}}
\def\epsilon{\varepsilon}
\def\degr{\hbox{$^\circ$}}
\def\arcmin{\hbox{$^\prime$}}
\def\arcsec{\hbox{$^{\prime\prime}$}}
\let\ts=\thinspace
\let\picplace=\vspace
\title*{Techniques of Radio Astronomy}
 \titlerunning{Techniques of Radio Astronomy} 
\author{T.~L.~Wilson\inst{1}}
\institute{Code 7210, Naval Research Laboratory, 4555 Overlook Ave., SW, Washington DC 20375-5320
\texttt{tom.wilson@nrl.navy.mil}}
\maketitle
\section*{Abstract} 
This chapter provides an overview of the techniques of radio astronomy. This study began in 1931 with Jansky's discovery of emission from the cosmos, but the period of rapid progress began fifteen years later. From then to the present, the wavelength range expanded from a few meters to the sub-millimeters, the angular resolution increased from degrees  to finer than milli arc seconds and the receiver sensitivities have improved by large  factors. Today, the technique of aperture synthesis produces images comparable to or exceeding those obtained with the best optical facilities. In addition to technical advances, the scientific discoveries made in the radio range have contributed much to opening new visions of our universe. There are numerous national radio  facilities  spread over the world. In the near future, a new era of truly global radio observatories will  begin. This chapter contains a short history of the development of the field, details of calibration procedures, coherent/heterodyne and incoherent/bolometer receiver systems, observing methods for single apertures and interferometers, and an overview of aperture synthesis. 
\vspace{5mm}

\noindent
{\it keywords}:  Radio Astronomy--Coherent Receivers--Heterodyne Receivers--Incoherent Receivers--Bolometers--Polarimeters--Spectrometers--High Angluar Resolution--Imaging--Aperture Synthesis

\section{Introduction}
Following a short introduction, the basics of simple  radiative transfer, propagation through the interstellar medium, polarization,  receivers, antennas, interferometry and aperture synthesis are presented. References are given mostly to more recent publications, where citations to earlier work can be found; no internal reports or web sites are cited.  The units follow the usage in the astronomy literature.  For more details, see Thompson et al.~(2001), Gurvits et al.~(2005), Wilson et al.~(2008), and  Burke \& Graham-Smith (2009).

\begin{figure}
\label{EM-spectrum}
\includegraphics[width=12.0cm,height=9.9cm]{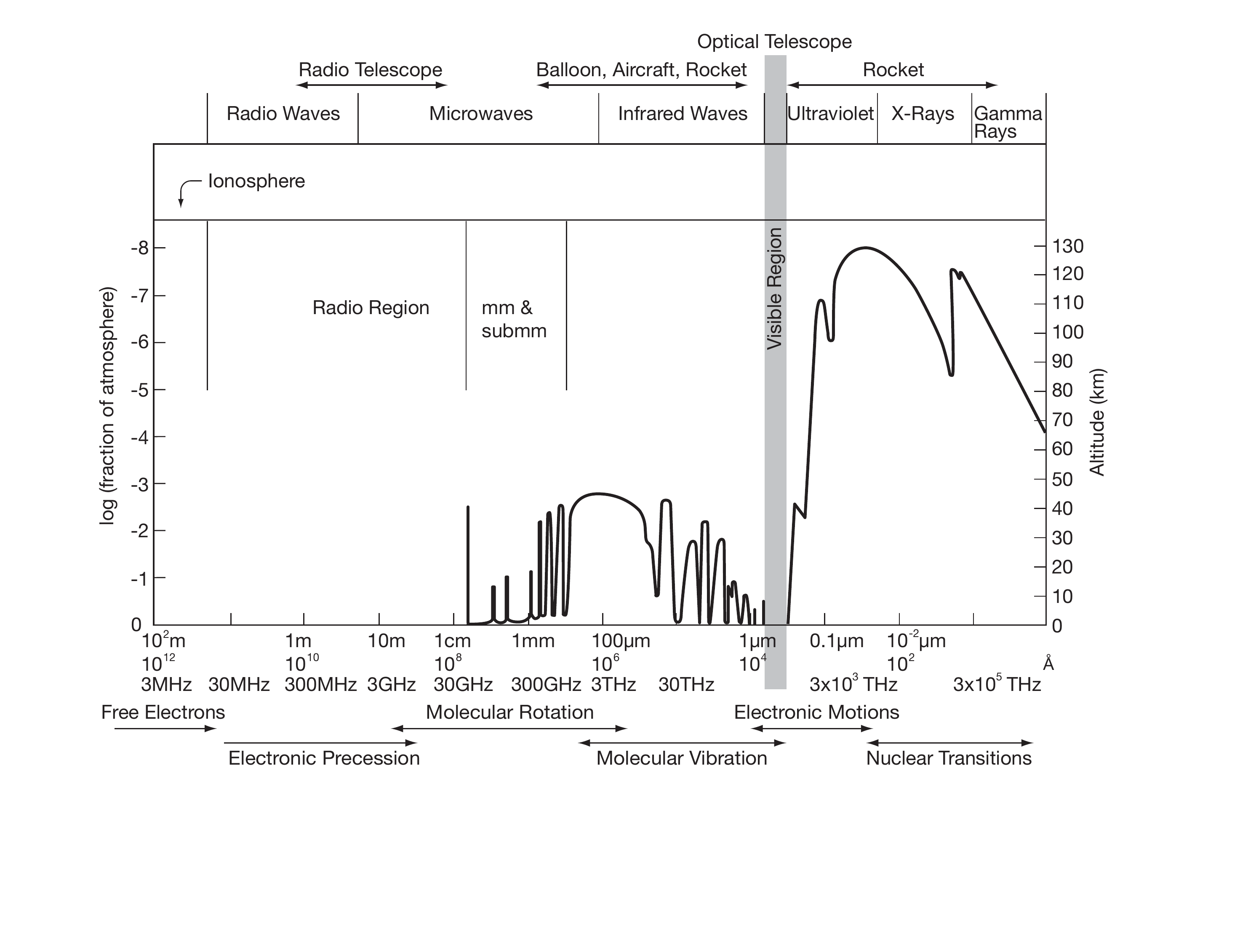}
  \caption[transmission through atmosphere versus wavelength]
 {  A plot of transmission through the atmosphere  versus wavelength, $\lambda$ in metric units and frequency, $\nu$,  in Hertz. 
The thick curve gives the fraction of the atmosphere  (left vertical axis) and the altitude (right axis) needed to reach a transmission of 0.5. The fine scale variations in the thick curve are caused by molecular transitions (see Townes \& Schawlow 1975). The thin vertical line on the left ($\sim10$MHz) marks the boundary where  ionospheric effects impede astronomical measurements. The labels above indicate the types of facilities needed to measure at the 
frequencies and wavelengths shown. For example, from the thick curve, at $\lambda$=100 $\mu$m,  one half  of the astronomical signal penetrates to an altitude  of 45 km.  In contrast,  at $\lambda$=10 cm, all of this signal is present at  the earth's surface. The arrows at the bottom of the figure indicate the type of atomic or nuclear process that gives rise to the radiation at the frequencies and wavelengths shown above (from Wilson et al.~2008). 
  }
\end{figure}

The origins of optical astronomy are lost in pre-history. In contrast radio astronomy began recently, in 1931, when K. Jansky 
showed that the source of excess radiation at $\nu=$20.5 MHz ($\lambda=$14.6 m) arose from outside the solar system. G. Reber followed up and extended Jansky's work, but the most rapid progress occurred after 1945,  when the field   developed quickly. The studies included broadband radio emission from the Sun, as well as emission from extended regions 
in our galaxy, and later other galaxies. In wavelength, the studies began at a few meters  where the emission was rather intense and more easily measured (see Sullivan 2005, 2009). 
Later, this was   expanded to include centimeter, millimeter and then sub-mm wavelengths. In Fig.~\ref{EM-spectrum} a plot of  transmission through the atmosphere as a function of frequency $\nu$ and wavelength, $\lambda$ is presented.  The extreme limits of the earth-bound radio window extend roughly from a lower
frequency  of $\nu \cong 10$\,MHz ($\lambda \cong 30$\,m) where the ionosphere sets a limit, to a highest frequency
of $\nu \cong 1.5$\,THz ($\lambda \cong 0.2  $\,mm), where molecular transitions of atmospheric  H$_2$O and N$_2$ absorb astronomical signals. There is also a prominent atmospheric feature at $\sim55$ GHz,  or 6 mm, from O$_2$.  The limits shown in 
Fig.~\ref{EM-spectrum} are not sharp since there are variations
both with altitude, geographic position and  time.  Reliable measurements at the shortest wavelengths   
 require remarkable sites on earth. Measurements at wavelengths  shorter than $\lambda$=0.2 mm  require the use of high flying aircraft, balloons or satellites. The curve in Fig.~\ref{EM-spectrum} allows  an estimate of the 
height above sea level needed to carry out astronomical measurements. 
%

The broadband emission mechanism that dominates at meter wavelengths has been associated with the synchrotron process. Thus although the photons have 
energies in the micro electron volt  range, this  emission is caused by  highly relativistic  electrons (with $\gamma$ factors of more than 10$^3$) 
moving in microgauss fields. In the centimeter and
millimeter wavelength ranges, some broadband emission is produced by the synchrotron process, but additional emission arises from free-free Bremsstrahlung from ionized gas near high mass stars and quasi-thermal broadband emission from dust grains. 
In the mm/sub-mm range,  emission from dust grains dominates, although free-free and synchrotron emission may also contribute. Spectral lines of molecules become more prominent at mm/sub-mm wavelengths  (see Rybicki \& Lightman 1979, Lequeux 2004, Tielens 2005). 

Radio astronomy measurements are carried out at wavelengths vastly longer than those used in the optical range  (see Fig.~\ref{EM-spectrum}), so  extinction of radio waves by dust is not an important effect. However, the longer wavelengths lead to lower angular resolution, $\theta$, since this is proportional
to $\lambda$/D where D is the size of the aperture  (see Jenkins \& White 2001). In the 1940's, the angular resolutions of radio telescopes were on scales of many arc minutes, at best. In time, 
interferometric techniques were applied to radio astronomy, following the method first used by  Michelson. This was further developed, resulting in  Aperture Synthesis, mainly by M. Ryle and associates at Cambridge University  (for a history, see Kellermann \& Moran 2001).  Aperture synthesis has allowed imaging with angular resolutions finer than milli arc seconds with facilities such as the Very Long Baseline Array (VLBA).  

Ground based measurements  in the sub-mm wavelength range have been made possible by the erection of facilities on 
extreme sites such as  Mauna Kea, the South Pole and the 5 km high site  of the  Atacama Large Millimeter/sub-mm Array (ALMA). Recently there has been renewed interest in high resolution imaging at  meter wavelengths. This is  due to the use of corrections for  smearing  by fluctuations in the electron content of the  ionosphere and advances that facilitate imaging over wide angles (see, e.g.,~Venkata 2010). With time, the general trend has been toward higher sensitivity, shorter wavelength, and higher angular resolution. 

Improvements in angular resolution have been accompanied by improvements in receiver sensitivity. Jansky used the highest quality receiver system then available. Reber had access to excellent systems. At the longest wavelengths, emission from  astronomical sources   dominates. At mm/sub-mm wavelengths, the transparency of the earth's atmosphere is an important factor, adding both noise and attenuating the astronomical signal, so both lowering receiver noise and measuring from high, dry sites  are  important. At meter and cm wavelengths, the sky is more transparent and radio sources are weaker.

The history of radio astronomy is replete with major discoveries. The first was implicit in the data taken by Jansky. In this, the intensity of the extended radiation from the Milky Way exceeded that of the quiet Sun. 
This  remarkable fact  shows that radio and optical  measurements sample fundamentally different phenomena. The radiation measured by Jansky was caused by the synchrotron mechanism; this interpretation was made more than 15 years  later (see Rybicki \& Lightman 1979). The next discovery, in the 1940's, showed that the active Sun caused disturbances seen in radar receivers. In Australia, a unique instrument was used to associate this variable emission with sunspots (see Dulk 1985, Gary \& Keller 2004). Among  later discoveries have been: (1) discrete cosmic radio sources, at first, supernova remnants and radio galaxies (in 1948, see Kirshner 2004), (2)  the 21 cm line of atomic hydrogen  (in 1951, see Sparke \& Gallagher 2007, Kalberla et al. 2005), (3) Quasi Stellar Objects (in 1963, see Begelman \& Rees 2009), (4) the Cosmic Microwave Background (in 1965, see Silk 2008), (5) Interstellar molecules (see Herbst \& Dishoeck 2009) and 
the connection with Star Formation, later including circumstellar and protoplanetary disks (in 1968, see Stahler \& Palla 2005, Reipurth et al.~2007),  
(6) Pulsars (in 1968, see Lyne \& Graham-Smith 2006), (7) distance determinations using source proper motions determined from Very Long Baseline Interferometry (see Reid 1993) and (8) molecules in high redshift sources (see Solomon \& Vanden Bout 2005). These areas of research have led to investigations such as the dynamics of galaxies,  dark matter, tests of general relativity, Black Holes, the  early universe and  gravitational radiation (for overviews see Longair 2006, Harwit 2006). Radio astronomy has been recognized by the physics community in that four Nobel Prizes (1974, 1978, 1993 and 2006) were awarded for work in this field. In chemistry, the community has been made aware of the importance of a more general chemistry involving ions and molecules (see Herbst 2001). Two Nobel Prizes for chemistry were awarded to persons actively engaged in molecular line astronomy.

Over time, the trend   has been away  from small groups of researchers constructing special purpose instruments toward the establishment of large facilities where users propose projects carried out by specialized staffs. These large facilities are in the process of becoming global. Similarly, the evolution of data reduction has been toward standardized packages developed by large teams. In addition, the demands of the interpretation of astronomical phenomena  have led to  multi-wavelength  analyses interpreted with the use of detailed models. 

 Outside the norm are projects designed to measure a particular phenomenon. A prime example is the study of the cosmic microwave background\index{Cosmic Microwave Background} (CMB) emission from the early universe. CMB data were taken with the COBE and WMAP satellites. These results showed that the CMB is is a Black Body (see Eq.~\ref{Planck-expression}) with a temperature of 2.73~K. Aside from a dipole moment caused by our motion, there is angular structure in the CMB at a very low level; this is being studied with the PLANCK satellite. Much effort continues to be devoted to measurements of the polarization of the CMB with ground-based experiments such as BICEP,  CBI, DASI and QUIET. For details and references to other CMB experiments, see their websites. In spectroscopy, there have been extensive surveys of the 21 cm line of atomic  hydrogen, H~I  (see Kalberla et al.~2005) and the rotational $J=1-0$ line from the ground state of carbon monoxide (see Dame et al.~1987).  These surveys have been extended to external galaxies (see Giovanelli \& Haynes 1991). During the Era of Reionization (redshift $z\sim$10 to 15), the H~I  line is  shifted to meter wavelengths. The detection of such a feature is the goal of a number of individual groups, under the name HERA (Hydrogen Epoch of Reionization Arrays). 

\subsection{A Selected List of Radio Astronomy Facilities}
 There are a large number  of existing  facilities; a selection is listed here.
General purpose instruments include the largest single dishes: the Parkes 64-m, the Robert C. Byrd Green Bank Telescope, hereafter GBT, the Effelsberg 100 meter, the 15-m James Clerk Maxwell Telescope (JCMT), the IRAM 30-m millimeter telescope and the 305-m Arecibo instrument. All of these have been in operation for a number of years. Interferometers form another category of instruments. The Expanded Very Large Array, the EVLA, is now in the test phase with $''$shared risk$''$ observing. Other large  interferometer systems are the VLBA, the Westerbork Synthesis Radio Telescope in the Netherlands, the Australia Telescope, the Giant Meter Wave Telescope in India, the MERLIN array a number of  arrays at Cambridge University in the UK and the MOST facility in Australia. In the mm range,  CARMA  in California and Plateau de Bure in France are in full operation, as is the Sub-Millimeter Array of the Harvard-Smithsonian CfA and ASIAA on Mauna Kea, Hawaii. At longer wavelengths, the Low Frequency Array, LOFAR, has started the first measurements and will expand by adding stations throughout Europe. The Square Kilometer Array, the SKA, is in the planning phase as is the FASR solar facility, while the Australian SKA Precursor  (ASKAP), the South African SKA precuror, (MeerKAT), the Murchison Widefield Array in Western Australia and Long Wavelength Array in New Mexico  are under construction. A portion of the Allen Telescope Array, ATA,  is in operation. A number of facilities are under construction, being commissioned or have recently become operational. At sub-mm wavelengths, the Herschel Satellite Observatory has been delivering data. The Five Hundred Meter Aperture Spherical Telescope, FAST, a design  based on the Arecibo instrument, is being planned  in China. This will be the world's largest single aperture.  The Large Millimeter Telescope, LMT, a joint Mexican-US project, will soon begin science operations as will the Stratospheric Far-Infrared Observatory (SOFIA) operated by NASA and the German DLR organization. Descriptions of these instruments are to be found in the internet.  Finally, the most ambitious ground based astronomy project to date is ALMA  which will start early science operations in late 2011 (for an account of  the variety of ALMA science goals, see Bachiller \& Cernicharo 2008). 

\section{Radiative Transfer and Black Body Radiation}
    \label{sbd}

 The total\index{Flux!total} flux of a source is obtained by
integrating Intensity (in Watts m$^{-2}$ Hz$^{-1}$ steradian$^{-1}$) 
over the total solid angle $\Omega_{\rm s}$ subtended by the source
\begin{equation}
 \label{total-flux}
  S_\nu = \int\limits_{\Omega_{\rm s}} I_\nu (\theta,\varphi)
                   \cos \theta \, {\rm d}\Omega .
\end{equation}
The flux density of astronomical sources is given in 
 units of the Jansky (hereafter Jy), that is, 
  $1\,{\rm Jy} = 10^{-26} \,{\rm W \, m}^{-2} {\rm Hz}^{-1}$.

The {\em equation of transfer\index{Transfer!equation}} is useful in interpreting the behavior of astronomcial sources, receiver systems, the effect of the earth's atmosphere on measurements.  
Much of this analysis is based on a one dimensional version of the general expression as  (see Lequeux 2004 or Tielens 2005): 
\begin{equation}
 \label{eq-of-transfer}
 \fboxsep2mm
 \fbox{$ \displaystyle
 \frac{{\rm d}I_\nu}{{\rm d}s} = - \kappa_\nu I_\nu + \epsilon_\nu $}
\quad .
\end{equation}
 The linear absorption coefficient  $\kappa_\nu$ and the emissivity $\epsilon_\nu$ are independent of the
intensity $I_\nu$. From the {\it optical\index{Depth!optical} depth} 
 definition ${\rm d}\tau_\nu = - \kappa_\nu \,{\rm d}s$,  the Kirchhoff relation $\epsilon_\nu/\kappa_\nu=B_\nu$ (see (Eq.~\ref{Planck-expression}))
and the assumption of an isothermal medium, the result is:

 \begin{equation}
  \label{sol-isoth}
   \fboxsep2mm
   \fbox{$ \displaystyle
     I_\nu(s) = I_\nu(0) \,{\rm e}^{- \tau_\nu(s)} + B_\nu(T)\,
                (1 - \,{\rm e}^{- \tau_\nu(s)}) $} \quad .
  \end{equation}
For a large optical\index{Optical depth} depth, that is for $\tau_\nu(0)
\rightarrow \infty$, (Eq.~\ref{sol-isoth}) approaches the limit
 \begin{equation}
  \label{large-opt-d}
   I_\nu = B_\nu(T)\ts .
 \end{equation}
This is case for planets and the 2.73~ K CMB. 
From  (Eq,~\ref{sol-isoth}), the difference between $I_\nu(s)$ and  $I_\nu(0)$ gives
\begin{equation}
 \label{I-comp}
 \Delta I_\nu(s)=I_\nu(s) - I_\nu(0) =
   (B_\nu(T)-I_\nu(0))(1-\expo^{-\tau}) \; .
\end{equation}
this represents the result of  an on-source minus off-source measurement, which is relevant for discrete sources.

The spectral distribution of the radiation of a black\index{Black body!radiation} body in thermodynamic equilibrium is given by the
Planck\index{Planck function} law 
\begin{equation}
\label{Planck-expression}
   \fboxsep2mm
   \fbox{$ \displaystyle
     B_\nu(T) = \frac{2 h \nu^3}{c^2} \frac{1}{{\rm e}^{h\nu / kT} -1}
        $} \quad .
\end{equation}
If $ h \nu \ll k T$, the {\em Rayleigh\index{Rayleigh-Jeans law}-Jeans Law} is obtained:  
\enlargethispage{2\baselineskip}%
\noindent
 \begin{equation}
  \label{b-jeans}
   \fboxsep2mm
   \fbox{$ \displaystyle
    B_{\rm RJ}(\nu, T) = \frac{2 \nu^2}{c^2} k T $} \quad .
  \end{equation}
 %

In the Rayleigh-Jeans relation,  the brightness and the thermodynamic temperatures of 
Black Body  emitters are strictly proportional
(Eq.~\ref{b-jeans}). This feature is useful, so the normal  
 expression of brightness of an extended source 
is {\em brightness temperature\index{Temperature!brightness}} $\tb$: 
 \begin{equation}
  \label{t-b}
    \tb = \frac{c^2}{2 k} \frac{1}{\nu^2}\, I_\nu =
   \frac{\lambda^2}{2 k} \, I_\nu \ts .
 \end{equation}


If $I_\nu$ is emitted by a black body and $h \nu \ll k T$ then
(Eq.~\ref{t-b}) gives the thermodynamic temperature of the source, a
value that is independent of $\nu$. If other processes are responsible for the emission of the
 radiation (e.g.,~synchrotron, free-free or broadband dust emission), $T_{\rm B}$ will depend on the
 frequency; however    (Eq.~\ref{t-b}) is still   used. If  the condition $\nu(\rm GHz) \ll 20.84 \left( T(\rm K) \right) $  is not valid, (Eq.~\ref{t-b}) can still be applied,
but  $T_{\rm B}$ will differ 
from the thermodynamic temperature of a black body. However, corrections are  
simple to obtain.

If (Eq.~\ref{t-b}) is combined with  (Eq.~\ref{I-comp}), the result is an expression for brightness 
temperature: 
\beq
 J(T)=\frac{c^2}{2 k \nu^2}(B_\nu(T)-I_\nu(0))(1-\expo^{-\tau_\nu(s)})
\; .
\eeq
The expression 
$J(T)$ can be expressed as  a temperature in most cases. This quantity is referred to as $T^*_{\rm R}$,
the {\em radiation temperature} in the mm/sub-mm range, or the {\em brightness temperature},
$T_{\rm B}$ for longer wavelengths. In the Rayleigh-Jeans approximation the equation of transfer is: 

 \begin{equation}
  \label{eq-tb}
   \fboxsep2mm
   \fbox{$ \displaystyle
       \frac{{\rm d}T_{\rm B}(s)}{{\rm d}\tau_\nu} =
           T_{\rm bk}(0) - T(s) $} \quad ,
 \end{equation}
where $T_{\rm B}$ is the measured quantity, $T_{\rm bk}(s)$ is the background source temperature and $T(s)$ is the temperature of the intervening medium
If the medium is isothermal\index{Isothermal!medium}, the general (one dimensional) solution  becomes
 \begin{equation}
  \label{tb-isoth}
   \fboxsep2mm
   \fbox{$ \displaystyle
     T_{\rm B} = T_{\rm bk}(0) \,{\rm e}^{- \tau_\nu(s)} +
          T \, (1- \,{\rm e}^{- \tau_\nu(s)})   $} \quad .
  \end{equation}

 \subsection{The Nyquist\index{Nyquist theorem} Theorem and  Noise Temperature}

This theorem relates the thermodynamic quantity temperature to the electrical quantities voltage and power. This is essential for the analysis of noise in receiver systems. 
The average power per unit bandwidth, $ P_\nu$  (also referred to as  Power Spectral Density, PSD), produced by a  resistor $R$  is
\begin{equation}
 \label{pow}
 P_\nu = \langle iv \rangle = \frac{\langle v^2 \rangle}{2R} =
  \frac{1}{4R}\langle v_{\rm N}^2 \rangle\ts ,
\end{equation}
where $v(t)$ is the voltage that is produced by $i$ across $R$, and
$\langle\cdots\rangle$ indicates a time average. The first factor $\frac{1}{2}$ arises from the condition for the  transfer of maximum power from $R$ over a  broad 
range of frequencies. The second factor
$\frac{1}{2}$ arises from the time average of $v^2$.  
 %
Then 
\begin{equation}
 \langle v_{\rm N}^2 \rangle = 4 R \,k \,T \; .
\end{equation}
When inserted  into (Eq.~\ref{pow}), the result is 
\begin{equation}
\label{Nyquist}
 P_\nu = k\, T  \; .
\end{equation}
(Eq.~\ref{Nyquist}) can also be obtained by a reformulation of the Planck law for one dimension in the Rayleigh-Jeans limit. Thus, the available noise power of a
resistor is proportional to its temperature, the {\em
noise\index{Temperature!noise} temperature} $T_{\rm N}$,  independent
of the value of $R$ and of  frequency.  
 
Not all circuit elements can be characterized by thermal noise. For
example a microwave oscillator can deliver 1 $\mu$W, the equivalent of more than 
$10^{16}$\,K, although the physical
temperature is $\sim$300\,K. This is an example of a very {\em nonthermal} process, so temperature 
is not a useful concept in this case.

\subsection{Overview of Intensity, Flux Density and Main Beam Brightness Temperature}
Temperatures in radio astronomy have given rise to some confusion.  A short summary with references to later sections is given here. Power is measured by an instrument consisting of an antenna and receiver. The power input can be calibrated and expressed as Flux Density or Intensity. 
For very extended sources, Intensity (see (Eq.~\ref{t-b})) can be expressed as a temperature, the  {\em main beam brightness temperature}, T$_{\rm MB}$. To obtain T$_{\rm MB}$, the measurements  must be  calibrated (Section~\ref{sd-cals}) and corrected  using the appropriate efficiencies (see Eq.~\ref{mbeam-eff} and following). For discrete sources,  the combination of   (Eq.~\ref{total-flux}) with (Eq.~\ref{t-b})  gives: 
\begin{equation}
 \label{flux-den-tb1}
    \fboxsep2mm
    \fbox{$ \displaystyle
S_{\nu}=\frac{2\, k \, \nu^2}{c^2} \tb \, \uDelta  \Omega  
        $}  \quad .
\end{equation}
For a  source with a Gaussian spatial distribution, this relation is 
 \begin{equation}
   \label{flux-den-tb2}
    \left[ \frac{S_{\nu}}{\rm Jy} \right] = 0.0736 \,  \tb \,  \left[ \frac{\theta}{\rm arc \, seconds} \right]^{2}  \left[  \frac{\lambda}{\rm mm} \right]^{-2}
          \end{equation}
 if the flux density $S_\nu$ and  the actual (or $''$true$''$) source size are known, then the {\em true brightness temperature}, $\tb$, of the source can be determined. For Local Thermodynamic Equilibrium (LTE),  $\tb$ represents the physical temperature of the source. If  the {\em apparent} source size, that is, the source angular size as measured with an antenna is known, (Eq.~\ref{flux-den-tb2})   allows a  calculation of  T$_{\rm MB}$.  For discrete sources, T$_{\rm MB}$  depends on the angular resolution.  If the antenna beam size (see Fig.~\ref{polar-pattern} and discussion)   has a Gaussian shape  $\theta_{\rm b}$, the relation of actual $\theta_{\rm s}$ and apparent size $\theta_{\rm o}$ is:
     \begin{equation}
     \label{gausssize0}
     \theta_{\rm o}^2 = \theta_{\rm s}^2 + \theta_{\rm b}^2 \, .
     \end{equation}
then from (Eq.~\ref{flux-den-tb1}), the relation of  T$_{\rm MB}$  and T$_{\rm B}$  is:
 \begin{equation}
     \label{gausssize1}
 T_{\rm MB}    \left( \theta_{\rm s}^2 + \theta_{\rm b}^2 \right) = T_{\rm B} \,  \theta_{\rm s}^2
     \end{equation}
%
 Finally, the PSD entering the receiver (Eq.~\ref{Nyquist}) is antenna temperature, T$_{\rm A}$; this is relevant for estimating signal to noise ratios (see (Eq.~\ref{aper-eff})  and  (Eq.~\ref{a-temperature})).  To establish  temperature scales and  relate received power to source parameters for filled apertures, see  Section~\ref{sd-cals}. For interferometry  and Aperture Synthesis, see  Section \ref{interferometers}. 

\subsection{Interstellar Dispersion and Polarization }

Pulsars\index{Pulsars}  emit radiation in a short time interval (see Lorimer \& Kramer 2004,  Lyne \& Graham-Smith 2006). If all frequencies are emitted at  
the same instant, the arrival time delay of different frequencies is caused by the ionized Interstellar Medium (ISM). This is characterized by the quantity $ \int_0^L N(l) \diff l$, which  is the column density of the
electrons to a distance $L$.
Since distances in astronomy are measured in parsecs  it has become customary to express 
the {\em dispersion\index{Dispersion measure} measure} as: 
\begin{equation}
\label{disp-measure-si}
{\mathrm{DM}} = \int\limits_0^L \left( \frac{N}{\mathrm{cm}^{-3}} \right)
\diff \left( {\frac{l}{\mathrm{pc}}} \right) 
\end{equation}
 The lower frequencies are delayed more in the ISM, so the relative time delay is: 
\begin{equation}
\label{delay-DMcm}
\displaystyle
\frac{\Delta \tau_{\mathrm{D}}}{\mug \rm s} = 1.34 \times 10^{-9}
\left[ \frac{\mathrm{DM}}{\mathrm{cm}^{-2}} \right]
\left[ \displaystyle \frac{1}{\left(
\displaystyle
\frac{\nu_1}{\mathrm{MHz}} \right)^2} - \frac{1}{\left(
\displaystyle
\frac{\nu_2}{\mathrm{MHz}} \right)^2} \right]
\end{equation}
Since both  time delay\index{Pulse!delay} $\Delta \tau_{\mathrm{D}}$ and 
observing
frequencies $\nu_1  < \nu_2$
can be measured with high precision, a very
accurate value of DM for a given pulsar can be determined. 
Provided the distance 
 to the pulsar, $L$, is known,  a good estimate of the
average electron density between observer and pulsar can be found. However since
$L$ is usually known with moderate accuracy, only approximate values
for $N$ can be obtained. Often the opposite procedure is
used: From reasonable values for $N$, a measured DM
provides information on the unknown distance $L$ to the pulsar.

Broadband linear polarization is caused  by  non-thermal processes  (see Rybicki \& Lightman 1979)  including Pulsar radiation, quasi-thermal emission from aligned, non-spherical dust grains\index{Polarization} (see Hildebrand 1983) and scattering from free electrons. Faraday rotation will change the position angle of linear polarization as the radiation passes through an ionized medium; this varies as  $\lambda^2$, so this effect is larger for longer wavelengths. It is usual to characterize polarization by  the four Stokes Parameters, which are the sum or difference of measured quantities.  The total intensity of a wave is given by the  parameter $I$. The amount and angle of linear polarization by the parameters $Q$ and $U$, while the amount and sense of circular polarization is given by the parameter $V$.  Hertz dipoles are sensitive  to a single linear polarization. By rotating the dipole over an angle perpendicular to the direction of the radiation, it is possible to determine the amount and angle of  linearly polarized radiation. Helical antennas or arrangements of two Hertz dipoles are sensitive to circular polarization.  Generally, polarized radiation  is  a combination of linear and circular, and is usually less than 100\% polarized, so  four Stokes parameters must be specified. The definition of the sense of circular polarization in radio astronomy is the same as in Electrical Engineering but {\em opposite} to  that used in the optical range; see Born \& Wolf (1965) for a complete analysis of polarization, using  the {\em optical} definition of circular polarization.
  Poincar\'e introduced a representation that permits an easy
visualization of all the different states of polarization of a vector wave. See Thompson et al.~(2001), Crutcher (2008), Thum et al.~(2008) or Wilson et al.~(2008) for more details. 

\section[Receiver Systems]
{Receiver Systems}
\sectionmark{Receivers}
 \label{stoch-proc}
\subsection{Coherent and Incoherent Receivers}
 Receivers are assumed to be linear power measuring devices, i.~e.~any  non-linearity  is a small quantity. There are two types of receivers: coherent and incoherent.
Coherent  receivers are those which preserve the phase of the input radiation while incoherent do not.   Heterodyne  (technically  $''$superheterodyne$''$\index{receiver!heterodyne})   receivers are those which those which shift the frequency of the input but preserve phase.  The most commonly  used coherent receivers employ  heterodyning, that is, frequency shifting (see Section \ref{coh-rx}). 
  The most commonly used incoherent receivers  are bolometers (Section \ref{bolo-rx}); these are direct detection receivers, that is,  operate at sky frequency.  Both coherent and incoherent receivers add noise to the astronomical input signal; it is assumed that the noise of both  the input signal and the receiver follow  Gaussian distributions.   The noise contribution of coherent receivers is expressed  in Kelvins. Bolometer noise is characterized by the {\em Noise Equivalent Power}, or NEP,  in units of Watts Hz$^{-1/2}$ (see Section \ref{cal-proc} and  Section  \ref{mm-cal-proc}).  NEP is  the input power level  which doubles the output power.   More extensive discussions of receiver properties are given in  Rieke (2002) or Wilson et al.~(2008).

To analyze the performance of a
receiver, the commonly accepted model is an ideal receiver with no internal noise, but connected  to two noise sources, one
for the external noise (including the astronomical signal) and a second for the  receiver noise. To be useful,
receiver systems must increase the input power level.  
The noise contribution is  characterized by the {\em Noise Factor}, $F$. If the signal-to-noise ratio at the input is expressed as $(S_1/N_1)$ and at the output as  $(S_2/N_2)$,  the noise factor is: 

\begin{equation}
\label{Noise-fact}
 F = \frac{S_1/N_1}{S_2/N_2}   \; .
\end{equation}
A further step is to assume that the signal is amplified by a gain factor $G$ but otherwise unchanged. Then  $S_2 = G \, S_1$ and: 
 \begin{equation}
\label{Noise-fact1}
 F = \frac{N_2}{G \, N_1}   \; .
\end{equation}
For a direct detection system such as a bolometer, $G=1$. For coherent receivers, there must be a minimum noise contribution (see Section \ref{min-noise}), so $F>1$.  For coherent receivers $F$ is expressed in temperature units as $T_R$  using the relation
 \begin{equation}
\label{Noise-fact2}
 T_R  = (F - 1) \cdot 290  {\rm K}  \; .
\end{equation}

\subsubsection{Receiver Calibration }
 \label{cal-proc}

Heterodyne receiver noise performance is usually expressed in   degrees Kelvin.  In the calibration\index{Receiver!calibration procedure} process,
a  power scale (the PSD) is established at the receiver input.  This  is  measured 
in terms of the
noise\index{Temperature!noise} temperature. To calibrate a
receiver, the noise temperature increment $\Delta T$
at the receiver input must be  related  to a given measured receiver output
increment $\Delta z$ (this applies to coherent  receivers which have a wide dynamic range and a total power or $''$DC$''$ response). 
 Usually  resistive loads at two known 
(thermodynamic)
temperatures $T_{\rm L}$ and $T_{\rm H}$ are used. The receiver outputs
are $z_{\rm L}$ and $z_{\rm H} $, while $T_{\rm L}$ and $T_{\rm H}$ are the resistive loads at two temperatures. The relations are:
 \begin{eqnarray*}
    z_{\rm L} &=& (T_{\rm L} + T_{\rm R}) \, G \ts,   \\
    z_{\rm H} &=& (T_{\rm H} + T_{\rm R}) \, G \ts,
 \end{eqnarray*}
taking 
 \begin{equation}
   \label{yhl}
      y = z_{\rm H} / z_{\rm L} \ts .
 \end{equation}
the result is:
 \begin{equation}
    \label{tsrc}
    \fboxsep2mm\fbox{$ \displaystyle
      T_{\rm rx}= \frac{T_{\rm H} - T_{\rm L} \, y}{y - 1} $}
      \quad ,
 \end{equation}
This is known as the $''$y-factor$''$\index{Receiver!y-factor}; the procedure is a $''$hot-cold$''$ measurement. The determination of the  y factor  is 
calculated in  the Rayleigh-Jeans limit. 
Absorbers at  temperatures of $T_{\rm H}$ and $T_{\rm L}$   are used to produce the inputs. 
Often  these are chosen to be 
at the ambient temperature ($T_{\rm H} \cong
293$\,K or $20\degr$\,C) and at the temperature of liquid nitrogen
($T_{\rm L} \cong 78$\,K or $-195\degr$\,C). When receivers are installed on antennas,  such   $''$hot-cold$''$ calibrations  are  done only  infrequently.  As will be discussed in Section  \ref{m-cm-cal-proc}, in the cm and meter range, calibration signals are provided by noise diodes; from measurements of sources with known flux densities  intensity scales are established. Any atmospheric corrections are assumed to be  small at these wavelengths.   
As will be discussed in Section  \ref{mm-cal-proc},    in the mm/sub-mm wavelength range, from measurements of an ambient load (or two loads at different temperatures), combined with measurements of  emission from the atmosphere and models of the atmosphere,  estimates of atmospheric transmission are made.

 Bolometer performance  
is characterized by the {\em Noise Equivalent Power}, or NEP, given in units of Watts Hz$^{-1/2}$.   The expression for NEP can be related to a receiver noise temperature. For ground based bolometer systems, background noise dominates. For these, the background noise is given as T$_{\rm BG}$: 
\begin{equation}
  \label{nepph}
 \fboxsep2mm
    \fbox{$ \displaystyle
 {\rm NEP}= 2 \epsilon \,k \,T_{\rm BG} \,\sqrt{\Delta \nu}
 $} \quad .
\end{equation}
here  $\epsilon$ is the emissivity of the background and  $\Delta \nu$ is the bandwidth.  
Typical values for ground-based mm/sub-mm bolometers are $\epsilon = 0.5$,   T$_{\rm BG} = 300$ K  and
$\Delta \nu= 100$ GHz. For these values, 
 NEP$= 1.3 \times 10^{-15}$ \, Watts\,Hz$^{-1/2}$.  
With the collecting area of the IRAM  30\,m  or the  JCMT telescopes,  sources in the milli-Jansky (mJy) range can be measured. 

Usually bolometers are $''$A.~C.$''$ coupled, that is, the output responds to {\em differences} in the input power, so 
hot-cold measurements are not useful for characterizing bolometers. The response of bolometers is usually determined by measurements of sources with known flux densities, followed by measurements at, for example, elevations of 20$^{\rm o}$, 30$^{\rm o}$, 60$^{\rm o}$ and 90$^{\rm o}$  to determine the atmospheric transmission (see Section \ref{incoherent-cals}).



\subsubsection{Noise Uncertainties due to Random Processes}
\label{staelin} 

The  noise contributions from source, atmosphere, ground, telescope surface and receiver  are always additive: 
\begin{equation}
\label{nois-rx-tsys}
   T_{\rm sys} = \sum T_i
\end{equation}
From Gaussian statistics, the Root Mean Square, RMS, noise is given by the mean value divided by the square root of the number of samples. From the estimate that the number of samples is given by the product of receiver bandwidth multiplied by the integration time, the result is:

 \begin{equation}
   \label{dicke}
   \fboxsep2mm\fbox{$ \displaystyle
    \uDelta T_{\rm RMS} = \frac{T_{\rm sys}}{\displaystyle
  \sqrt{\uDelta \nu \, \tau}} $} \quad .
 \end{equation}
  A much more elaborate derivation is to be found in Chapter 4 of Rohlfs \& Wilson (2004), while a somewhat simpler account is in Wilson et al.~(2008).
 The calibration process in (Section \ref{cal-proc}) allows the  receiver noise to be expressed in degrees Kelvin. 
The relation of T$_{\rm sys}$ to T$_{\rm rx}$ is   $T_{\rm sys}=T_{\rm A} + T_{\rm rx}$, where $T_{\rm A} $ represents the power entering the receiver; at some wavelengths $T_{\rm A}$ will dominate $T_{\rm rx}$\index{System Noise}. In the mm/sub-mm range, use is made of  T$_{\rm sys}^*$, the system noise outside the atmosphere, since the attenuation of astronomical radiation is large. 
This will be presented in Section \ref{observations} and following.

\subsubsection{Receiver Stability}
  \label{rec-stab}
 
Sensitive receivers are  designed to achieve
a low value for $T_{\rm rx}$. Since the signals received are of exceedingly low
power,  receivers must also provide large receiver gains, $G$ (of order $10^{12}$), for sufficient output power. Thus  even very small gain instabilities
can dominate the thermal receiver noise. Since receiver
stability\index{Receiver!stability} considerations are   of prime importance,  comparison switching was necessary for early receivers (Dicke 1946). 
Great advances have been made in improving receiver stability since the 1960's so the need for rapid switching is lessened.  In the meter and cm wavelength range, the time between  reference measurements has increased.  However in the mm/sub-mm range, instabilities of the atmosphere play an important role; to insure that  noise decreases following  (Eq.\,\ref{dicke}), the effects of atmospheric and/or receiver instabilities must be eliminated. For single dish measurements, atmospheric  changes can be compensated for by rapidly  differencing   a measurement of the target source and a  reference.  Such {\em comparison or $''$Dicke$''$ switched} measurements are necessary for ground-based observations\index{Receiver!comparison switching}.   
If a typical procedure consists of using a total power receiver to measure on-source for 1/2  of the total  time, then   an  off-source comparison for 1/2 of the time  and  taking  the 
 difference of on-source minus off-source measurements,  the $\uDelta T_{\rm RMS}$ will be  a factor of 2 larger than the value given by  (Eq.~\ref{dicke}).

\section{Practical Aspects of Receivers}
This section concentrates on   receivers that are currently in use. For more details see Goldsmith (1988), Rieke (2002), or Wilson et al.~(2008).
\subsection{Bolometer Radiometers}\label{bolo-rx}

Bolometers\index{Bolometer} operate by  use of the effect
that the resistance, $R$, of a material varies with the temperature. In the 1970's,  the most sensitive  bolometers were semiconductor devices pioneered by F. Low. This is achieved when the bolometer element is cooled to very low temperatures.
When radiation is incident, the characteristics change, so this is a measure of the intensity of the
incident radiation.  Because this is a thermal effect, it  is independent of
the frequency and polarization of the radiation absorbed. Thus bolometers are intrinsically
broadband devices. It is possible to 
mount a polarization-sensitive device before  the
bolometer and thereby measure the direction and degree of
linear polarization.  
Also,  it is possible to 
carry out  spectroscopy, if frequency sensitive
elements, either filters, Michelson or Fabry-Perot interferometers,
are placed before the bolometer
element\index{Bolometer!Spectral Line}. Since these spectrometers operate at the sky frequency, the fractional resolution ($\uDelta \nu/\nu$) is at best $\sim 10^{-4}$. 
The data from each bolometer detector element (pixel) must be read out and then amplified.

For single dish (i.~e.~filled apertures) broadband continuum measurements at 
$\lambda <$ 2 mm,
multi-beam 
  bolometers  are the most common systems and  such 
systems can have a large number of beams. A promising new development in bolometer receivers is
{\em Transition Edge Sensors} referred to as  TES bolometers.
These superconducting devices may allow more than an order
of magnitude increase in sensitivity, if the bolometer is
not background limited. For 
bolometers used on  earth-bound telescopes,  
 the improvement with TES systems may be only 
$\sim$2--3 times more sensitive than the semiconductor  bolometers, but TES's will allow readouts from a much larger number of pixels.

A number of  large
bolometer arrays have produced  numerous publications: (1) MAMBO2 (MAx-Planck-Millimeter Bolometer)\index{Bolometer!MAMBO2} used on the IRAM 30-m telescope at 1.3 mm, (2) SCUBA  (Submillimeter Common User
Bolometer Array\index{Bolometer!SCUBA};  Holland et al.~1999)  on the JCMT, (3) the  LABOCA (LArge Bolometer CAmera)\index{Bolometer!LABOCA} array  on the APEX 12 meter telescope, (4) SHARC (Sub-mm High Angular Resolution Camera) 
on the Caltech Sub-mm Observatory 10-m  telescope and (5) MUSTANG (MUtiplexed Squid TES Array) on the GBT.  SCUBA  will be replaced  with
SCUBA-2 now being constructed at the U.~K.~Astronomy Technology Center, and there are plans to replace the MUSTANG  array  by MUSTANG-2, which is a larger TES system.

\subsection{Coherent Receivers}
 Usually, coherent receivers make use of  
heterodyning  to shift 
the signal input frequency  without changing  other properties of the input signal; in practice, this  is carried out by the use of 
mixers (Section \ref{mixer-rx}). The heterodyne process is   used in all branches of 
communications technology;  use of heterodyning allows measurements with unlimited spectral resolution. Although heterodyne receivers have a number of components, these systems have more flexibility than bolometers.

\subsubsection{ Noise Contributions in Coherent Receivers}
\label{coh-rx}
The noise generated in the first element dominates the system noise. The mathematical expression is given by the {\em Friis} relation \index{Friis formula} which accounts for the effect of 
cascaded amplifiers on the noise performance of a receiver\index{Noise!cascaded amplifiers}:
 \begin{equation}
   \label{ampst}
   \fboxsep2mm\fbox{$ \displaystyle
      T_{\rm S} = T_{\rm S1} + \frac{1}{G_1} T_{\rm S2} + \frac{1}{G_1 G_2}
           T_{\rm S3} + \dots + \frac{1}{G_1 G_2 \dots G_{n-1}} T_{{\rm S}n}
       $} \quad .
 \end{equation} 
Where $G_1$ is the gain of the first element, and $ T_{\rm S1}$ is the noise temperature of this element. 
For $\lambda >$3 mm ($\nu<115$GHz), the best cooled first elements, High Electron Mobility Transistors (HEMTs), typically have $G_1=10^3$ and $T_{\rm S1}=50$K; for 
$\lambda <$0.8 mm, the best cooled first elements, superconducting mixers, typically have $G_1 \le  1$, that is, a small loss, and $T_{\rm S1} \le 500$K. The stage following the mixer should have the lowest noise temperature and high gain. 

 \subsubsection{Mixers}\label{mixer-rx}
Mixers have been used in heterodyne receivers since Jansky's time. At first these were metal-oxide-semiconductor or {\em Schottky} mixers\index{Mixer!schottky}. 
 Mixers allow the signal frequency to be changed without altering the characteristics of the signal.  
In the mixing process,  the input signal is multiplied by  an intense monochromatic signal from a {\em local oscillator}, LO\index{Oscillator!local}. The frequency stability of the LO signal is maintained by a stabilization device in which the LO signal is compared with a stable input, in recent times, an atomic standard. These  phaselock loop systems produce  a pure, highly stable, monochromatic signal. 
 The mixer can be operated in the  Double Sideband (DSB) mode, in which  two sky frequencies, $''$signal$''$ and $''$image$''$\index{Image!frequency} at equal 
separations   from  the LO frequency (equal to the IF frequency)
are shifted into   intermediate (IF) frequency band. For spectral line measurements, usually one sideband is wanted, but the other not. DSB operation adds both noise and (usually) unwanted spectral lines; for spectral line measurements, single sideband (SSB) operation is preferred. In SSB operation, the unwanted sideband is suppressed, at the cost of more complexity.  In the sub-mm wavelength ranges, DSB mixers are still commonly used as the first stage of a receiver; in the mm and cm ranges, SSB operation is now the rule.

A significant improvement can be obtained if the mixer junction is operated
in the superconducting mode. The noise temperatures  and  LO  power requirements of superconducting mixers  are much lower than Schottky mixers.  
 Finally, the physical layout of such
devices is simpler since the mixer is a planar device, deposited on
a substrate by lithographic techniques.
SIS\index{Mixer!superconducting} mixers consist of a
superconducting layer, a thin insulating layer and another
superconducting layer (see Phillips \& Woody 1982).

 %
\begin{figure}
  \includegraphics[width=0.8\linewidth]{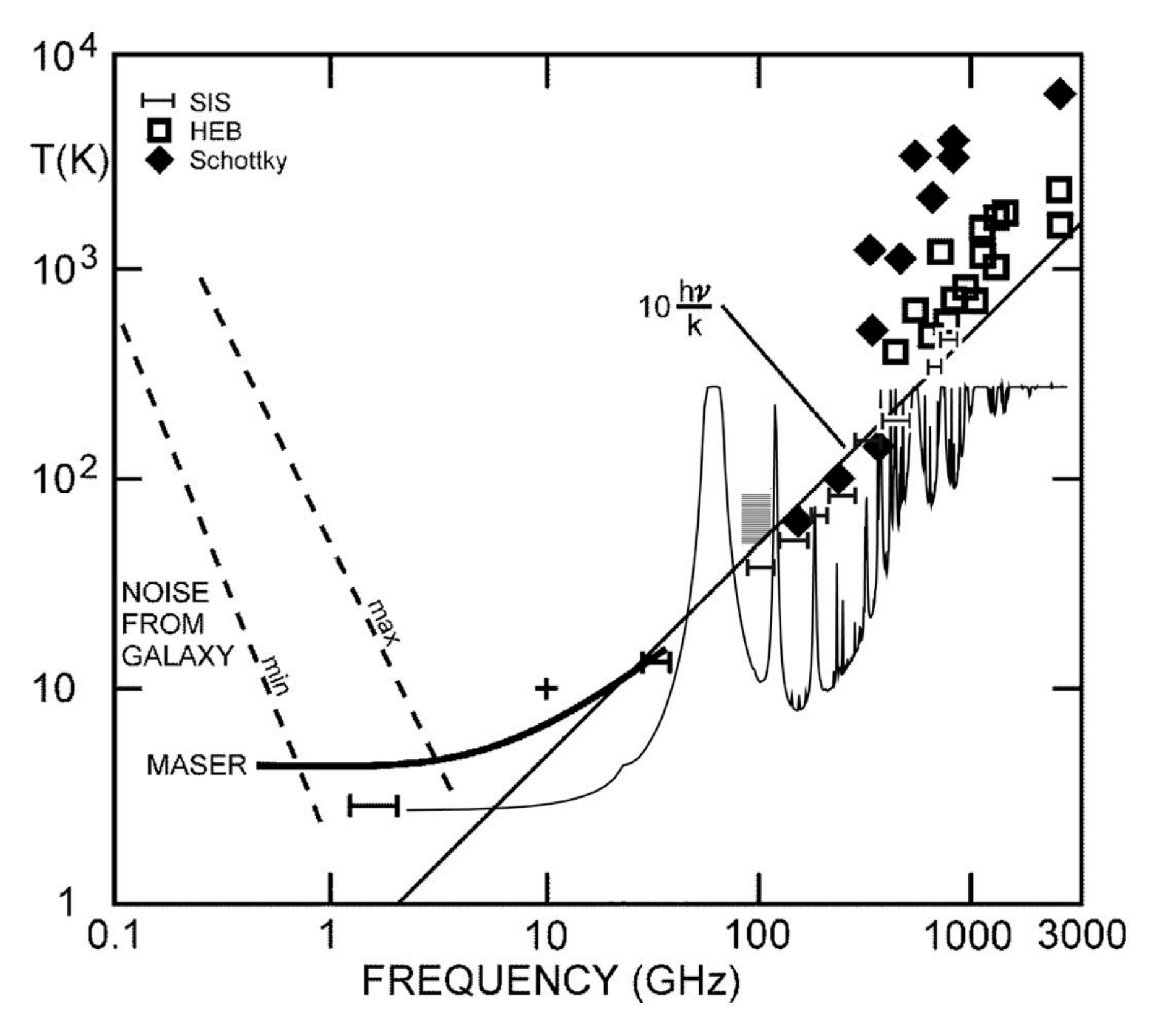}
  \caption[Receiver noise temperatures for  coherent
amplifiers]
 { \label{rntemp}
Receiver noise temperatures for coherent amplifier systems compared to
the temperatures from the Milky Way galaxy (at long wavelengths, on left part of figure) and the 
atmosphere (at mm/sub-mm wavelengths on the right side). The atmospheric emission is based on a model of zenith
emission for 0.4 mm of water vapor, that is, excellent weather (plot from B. Nicolic (Cambridge Univ.) using the $''$AM$''$ program of S. Paine (Harvard-Smithsonian Center for Astrophysics)). This does not take
into account the absorption of the astronomical signal. In the 1 to
26 GHz range, the two horizontal lines represent the noise temperatures of the best 
HEMT amplifiers, while the solid line represents the noise temperatures of maser receivers.  The shaded region between 85 and
115.6 GHz is the receiver noise for the SEQUOIA array (Five College
Radio Astronomy Observatory) which consists 
of monolithic millimeter integrated circuits (MMIC). The meaning of the other symbols is given
in the upper left of the diagram (SIS's are Superconductor-Insulator-Superconductor mixers, HEB's are Hot Electron Bolometer mixers).  The double sideband (DSB) mixer noise
temperatures were converted to single sideband (SSB) noise temperatures by doubling the receiver noise. The ALMA mixer 
noise temperatures  are SSB, as are the  HEMT values. The  line marked $''$10 h$\nu$/kT$''$ refers to the limit described in  (Eq.~\ref{heisenberg6}). Some data used in this diagram are taken from Rieke (2002). The figure is from Wilson et al. (2008)}
\end{figure}
Superconducting Hot Electron Bolometer-mixers (HEB) are heterodyne 
devices, in spite of the name. These mixers make use of superconducting thin
films which have sub-micron sizes (see Kawamura et al. 2002).

 A number of  multi-beam heterodyne cameras are in operation  in the cm range, but only a few in the mm/sub-mm range. 
The first mm multi-beam system was  the SEQUOIA array receiver pioneered by S. Weinreb; such devices are becoming more common. In contrast, 
multibeam systems that use SIS front ends are rare. Examples are a 9 beam Heterodyne Receiver Array of
SIS mixers at 1.3 mm, HERA,  on the IRAM 30-m
millimeter telescope, 
HARP-B, a 16 beam SIS system in operation at the JCMT for 0.8 mm and the CHAMP+ receiver at the Max-Planck-Inst. f\"ur Radioastronomy on the APEX 12-m telescope.

 \subsubsection{Square Law Detectors}

For heterodyne receivers  the input is normally amplified (for $\nu < 115$GHz), translated in frequency and then detected in a device 
that produces an output signal $y(t)$ which is proportional to the square of  $v(t)$:
 \begin{equation}
   \label{yax}
     y(t) = a \, v^2(t) 
 \end{equation}
Once detected, phase information is lost. For interferometers, the output of each antenna is a voltage, shifted in frequency and then digitized. This output is brought to  a central location for correlation.

 \subsubsection{The Minimum Noise in a Coherent System}
\label{min-noise}

     The ultimate limit for  coherent receivers or amplifiers is obtained
by an application of the {\em Heisenberg\index{Heisenberg uncertainty principle}
uncertainty principle} involving  phase and number of photons.  From this, the $minimum\index{Noise!minimum}$ noise of a coherent  amplifier 
     results in a receiver noise temperature of
\begin{equation}
     \label{heisenberg6}
       \fboxsep2mm\fbox{$ \displaystyle
            T_{\rm rx}({\rm minimum}) = \frac{h \nu}{ k}
        $}\quad .
     \end{equation}
     For {\em incoherent} detectors, such as bolometers, phase is not
     preserved, so this limit does {\em not} exist. In the 
     mm wavelength region, this noise temperature limit is
     quite small; at $\lambda$=2.6 mm ($\nu$=115 GHz), this limit  is 5.5 K. The value for the ALMA receiver in this range is about 5 to 6 times the minimum. A significant difference between  radio and optical regimes is that the minimum noise in the radio range is small, so that  the power from a single receiver can be amplified and then divided.  For example, for the EVLA, the voltage output of all 351 antenna pairs are combined with little or no loss in the signal-to-noise ratio. Another example is given in Section \ref{polarimeters}, where a radio polarimeter can produce all  four Stokes parameters from two inputs without a loss of the signal-to-noise ratio. 

\subsection[Back Ends]{Back Ends: Polarimeters \& Spectrometers}
\sectionmark{Back Ends}
\label{crp}
The term $''$Back End$''$ is used to specify the devices following the IF amplifiers.  Many different  back ends  have been designed for
specialized purposes such as  continuum, spectral or polarization measurements. 

For a single dish continuum correlation receiver, the (identical) receiver input is divided, amplified in two identical systems and then the outputs are multiplied. The gain fluctuations are uncorrelated but the signals are, so the effect on the output is the same as with a Dicke switched system, but with {\em no} time spent on a reference.

\subsubsection{Polarimeters}
\label{polarimeters}
A typical heterodyne dual polarization receiver consists of two identical systems, each sensitive to one of the two orthogonal  polarizations,  linear or circular. Both systems must be connected to the same local oscillator to insure that the phases  have a definite relation. Given this arrangement, a polarimeter can provide values of all four Stokes parameters simultaneously.   All Stokes parameters can also be measured using  a single receiver whose input  is  switched from one sense of polarization to the other, but then the integration time for each polarization will be halved.

\subsubsection{Spectrometers}
 \label{spectrom}

Spectrometers\index{Spectrometer}  analyze the spectral
information contained in the radiation field. To accomplish this, the spectrometer 
 must be  SSB\index{Receiver!SSB}  and the frequency resolution $\Delta
\nu$ is usually very good, sometimes in the kHz range. In addition, the time stability
must be high. 
If a resolution of $\Delta \nu$ is to be achieved for the spectrometer,
all those parts of the system that enter critically into the
frequency\index{Frequency!response} response have to be maintained to
better than $0.1 \, \Delta \nu$. For an overview of the current state of spectrometers,  see  Baker et al.~(2007).

Conceptually, the simplest spectrometer is composed of a set of $n$ adjacent 
filters, each with a  bandwidth $\Delta \nu$. Following each filter is a square-law detector
and integrator. For a finer resolution, another set of $n$ filters must be constructed.

Another approach to spectral analysis is to Fourier Transform (FT) the input,  $v(t)$,  to obtain $v(\nu)$  and then  square  $v(\nu)$   to obtain the 
Power Spectral Density\index{Density!spectral power}. The maximum bandwidth is limited by the sampling rate.  
From (another!) Nyquist  theorem, it is necessary to sample  at a rate equal to twice the bandwidth. In the simplest scheme,  for a bandwidth of 1 GHz, the sampling must occur at a rate of 2 GHz.  After sampling and Fourier Transform, the output is squared to produce power in an  $''$FX$''$ autocorrelator\index{autocorrelator!FX}.   For $10^3$ samples, each channel will have a 1 MHz resolution.

 For  $''$XF$''$ systems\index{autocorrelator!XF}, the input $v(t)$ is multiplied (the $''$X$''$) with a delayed signal $v(t -\tau)$ to obtain the autocorrelation\index{Function!autocorrelation} function 
$R(\tau)$. This is then Fourier Transformed to obtain the spectrum.  For $10^3$ samples, there will be  $10^3$  frequency channels. 
For an XF system the time delays are performed in a set of serial digital shift registers  
 with a sample delayed by a time $\tau$.  Autocorrelation can also  be carried out 
with the help of analog devices using a  series of cable delay lines; these can provide very large bandwidths.   The first  XF system for astronomy was a digital autocorrelator built by S. Weinreb in 1963.

The two significant advantages of digital  spectrometers are: (1)
flexibility and (2) a noise behavior that follows $1/\sqrt{t}$
after many hours of integration. The flexibility allows the choice of 
many different frequency resolutions and bandwidths or even to employ a
number of different spectrometers, each with different bandwidths,
simultaneously. 

A serious drawback of  digital auto and cross correlation
spectrometers had been limited bandwidths.  However, advances in digital technology  in recent years have allowed the construction of 
autocorrelation spectrometers with several 10$^3$ 
channels covering instantaneous   bandwidths of several GHz. 

 Autocorrelation systems are used in single antennas. The calculation of spectra makes use of
the symmetric nature
of the  autocorrelation function, ACF, so the number of delays gives the number of spectral channels. 

Cross-correlators  are used in interferometers and in some single dish applications. When used in an interferometer, the cross-correlation is between  different inputs so  will not necessarily be symmetric. 
 Thus,  the zero delay of the cross-correlator is placed  in channel $N/2$. The number of delays, $N$, allows the
determination of $N/2$ spectral intensities, and $N/2$ phases. The cross-correlation hardware can employ 
either an XF or a FX correlator.   For more details about the use of cross-correlation, see Section \ref{interferometers}. 


 
Until recently,  spectrometers with
bandwidths of several GHz often made use of Acoustic Optical analog techniques. 
    The Acoustic Optical Spectrometer (AOS) makes use of the diffraction of light by ultrasonic waves: these  cause
 periodic density variations in the crystal through which it passes.
These density variations in turn cause variations in the bulk
constants  of the crystal, so that a plane light 
 wave passing through this medium will be  modulated by the interaction with the crystal. The modulated light is detected in a charge coupled device. Typical AOS's have an instantaneous bandwidth of 2 GHz and 2000 spectral channels. 

In all cases, the spectra of the individual channels of a spectrometer  are expressed in terms of temperature with the relation:  

\begin{equation}
\label{correl-cal-base}
   T_i = \left [ \left( S_i - R_i \right)/R_i \right] \cdot T_{\rm sys}
\end{equation}
where $S_i$ is the normalized spectrum of channel $i$ for the on-source measurement and $R_i$ is the corresponding reference for this channel. For mm/sub-mm spectra, $T_{\rm sys}$ is replaced by $T_{\rm sys}^*$  (corrected for atmospheric losses; see Section \ref{mm-cal-proc}). For cross-correlators, as used in interferometers, the signals from two antennas are multiplied. In this case, the value of $T_{\rm sys}$ is the square root of the product of the system noise temperatures of the two systems.

\section{Antennas}
The antenna serves to focus power into  the feed, a device that efficiently transfers power in the electromagnetic wave  to the receiver\index{Antenna}.  According to the principle of  {\em reciprocity}\index{Antenna!reciprocity},  the properties of  antennas such as beam sizes, efficiencies etc. are the same whether these are 
 used for receiving or transmitting. Reciprocity holds  in astronomy, so it is usual to interchangeably use expressions that involve either transmission or reception when discussing  antenna properties.  All of the following applies to the far-field radiation.

\subsection{The Hertz Dipole }\label{hertz-larmor}
 The total power radiated from a Hertz dipole\index{Antenna!Hertz dipole} carrying an oscillating current $I$ at a wavelength $\lambda$ is 
 \begin{equation}
   \label{tot-rad-pow}
   \fboxsep2mm\fbox{$ \displaystyle
    P = \frac{2 c}{3}  \left(
                \frac{I \Delta l}{2 \lambda} \right)^2  $} \quad .
 \end{equation}
For the Hertz dipole, the radiation is linearly polarized with the electric field along the direction of the dipole. The radiation pattern has a donut shape, with the  
cylindrically symmetric maximum perpendicular to the 
axis of the dipole. Along the direction of the dipole, the radiation field is zero.  To improve directivity, reflecting screens have been placed behind a dipole, and in addition,  collections of  dipoles, driven in phase, are used.   Hertz dipole radiators have  the best efficiency when the size of the dipole  is $ 1/2 \, \lambda$  .

\subsection{Filled Apertures}
 \label{desap}
This Section  is a   simplified   description  of 
antenna properties 
 needed for the interpretation of  astronomical measurements. For more detail, see Baars (2007).  
 At cm and shorter wavelengths,  flared waveguides ($''$feed horns$''$)  or dipoles are used to convey power  focussed by the antenna  (i.~e., electromagnetic waves in free space) to the receiver (voltage). At the longest wavelengths, dipoles are used as the antennas.   
 Details are to be found  in Love (1976) and Goldsmith (1988, 1994).

 \subsubsection{Angular Resolution and Efficiencies}
From diffraction theory (see Jenkins \& White 2001), the angular resolution of a reflector of diameter $D$ at a wavelength $\lambda$ is 
 \begin{equation}
   \label{diffraction}
   \fboxsep2mm\fbox{$ \displaystyle
     \theta  = k\frac{\lambda}{D}  $}\quad .
\end{equation}
where $k$ is of order unity. This  universal result gives a value for $ \theta$ (here in radians when $D$ and $\lambda$ have the same units). Diffraction theory also predicts the unavoidable presence of sidelobes, i.~e. secondary maxima. The sidelobes can be reduced by {\em tapering} the antenna illumination. Tapering lowers the response to very compact sources and increases the value of $\theta$, i.~e. widens the beam. 
 
The reciprocity  concept provides a method to
measure the power pattern (response pattern or Point Spread Function, PSF\index{Antenna!point spread function})  using 
transmitters. However, the
distance from a large antenna A (diameter $D \gg \lambda$) to a  transmitter B (small in size)  must be so large that B produces plane waves across the aperture $D$ of antenna A, that is, so B is in the far field of A. 
This is the {\em Rayleigh} distance; it requires that the
curvature of a wavefront emitted by B is much less than $\lambda$/16  across the geometric dimensions of antenna  A.
By definition, at the Rayleigh distance $\mathcal{D}$,  the  curvature must be $ \gg D^2/8\lambda$\index{Antenna!Rayleigh distance} 
for an antenna of diameter $D$.

Often, the {\em normalized\index{Power!normalized pattern} power
pattern}  is measured:
 \begin{equation}
   \label{norm-pp}
   \fboxsep2mm\fbox{$ \displaystyle
     P_{\rm n} (\vartheta, \varphi) = \frac{1}{P_{\rm max}} \,P (\vartheta,
                 \varphi)  $}\quad .
\end{equation}
For larger apertures, the transmitter is usually replaced by  a small diameter
radio source of known flux density (see Baars et al.~1977, Ott et al.~1994). The flux densities of a few primary calibration sources are determined
by measurements using horn antennas at centimeter and millimeter wavelengths. At mm/sub-mm wavelengths, it is usual to employ 
planets, or moons of planets, whose surface temperatures are
known  (see Altenhoff 1985, Sandell 1994).

%
%
%
\begin{figure}
\sidecaption
    \includegraphics[width=2.6cm,height=5.9cm]{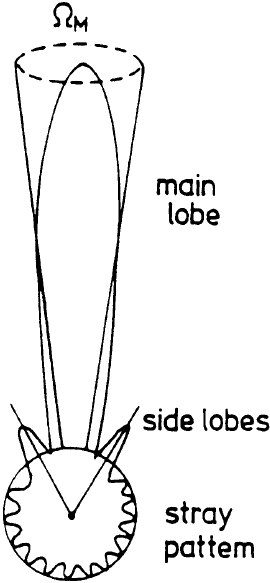}
  \caption[The polar power pattern]
 { \label{polar-pattern}
 A polar power pattern showing the main beam, and near and far
 sidelobes. The weaker far sidelobes have been combined to form
 the stray pattern}
\end{figure}
The {\em beam\index{Beam!solid angle} solid angle} $\Omega_{\rm A}$ of an
antenna is given by
 \begin{equation}
  \label{beam-sa}
  \fboxsep2mm
  \fbox{$ \displaystyle
  \Omega_{\rm A} = \parbox{8mm}
                       {$ \vspace*{-2mm}
                         {\displaystyle
                          \int\!\!\int \atop
                         {\!\!\!\!\!\scriptstyle 4 \pi}}  $  }
         P_{\rm n} (\vartheta, \varphi) \,\diff \Omega =
         \int\limits_0^{2\pi}\!\int\limits_{0}^{\pi}
        P_{\rm n} (\vartheta, \varphi) \sin \vartheta \diff \vartheta
          \diff \varphi $}
 \end{equation}
this is measured in steradians (sr). The integration is extended
over  all angles, so  $\Omega_{\rm A}$ is
the solid angle of an ideal antenna having $P_{\rm n} = 1$ for
 $\Omega_{\rm A}$ and $P_{\rm n} = 0$  everywhere else. For most antennas the (normalized) power pattern has
much larger values for a
limited range of both $\vartheta$ and $\varphi$ than for the remainder; the range where $\Omega_{\rm A}$  is large is  the
main beam  of the antenna; the remainder are  the sidelobes or backlobes
(Fig.\,\ref{polar-pattern}).  

In analogy to (Eq.~\ref{beam-sa})  the {\em
   main beam \index{Main beam}
solid angle} $\Omega_{\rm MB}$ is defined as 
 \begin{equation}
   \label{main-lsa}
     \fboxsep2mm\fbox{$\displaystyle
         \Omega_{\rm MB} =
                \mathop{\int\!\!\int}\limits_
               {\scriptstyle \rm main \atop
                \scriptstyle \rm  lobe}
              P_{\rm n} (\vartheta, \varphi) \,\diff \Omega
        $} \quad .
 \end{equation}
The quality of a single antenna  depends on how well the
power pattern is concentrated in the main beam. 
The 
definition of  {\em main beam efficiency\index{Beam!efficiency}} or 
{\em beam efficiency},
$\eta_{\rm B}$,  is:

 \begin{equation}
    \label{mbeam-eff}
    \fboxsep2mm\fbox{$ \displaystyle
       \eta_{\rm B} = \frac{\Omega_{\rm MB}}{\Omega_{\rm A}} $} \quad .
 \end{equation}%
 $\eta_{\rm B}$ is 
the fraction of
the power is concentrated in the
main beam. The main beam efficiency can be
modified (within  limits) for parabolic
antennas by changing the illumination  
of the main reflector.   An underilluminated antenna has a wider main beam but lower sidelobes. 
%
 The angular extent of the main beam is usually described
by the {\em full width to half power\index{Beam!width} width} (FWHP),
the angle between points of the main
beam where the normalized power pattern falls to $1/2$ of the maximum. 
For elliptically shaped  main beams, values for
widths in orthogonal directions are
needed. The beamwidth, $\theta$ is given by (Eq.~\ref{diffraction}).    If the FWHP beamwidth is well defined, the location of an isolated source is determined to the accuracy given by  the FWHP divided by  the S/N ratio. 
Thus, it is possible to determine positions to 
small fractions of the FWHP beamwidth, if the signal-to-noise ratio is high and noise is the only limit.

 %
If  a plane wave with the power density
$\mid\!\langle\vec{S}\rangle\!\mid$ in Watts m$^{-2}$
 is intercepted by an antenna, a certain
amount of power is  extracted  from
this wave. This 
power is $P_{\rm e}$ and the fraction is: 
\vspace{-6pt}
 \begin{equation}
   \label{a-e}
     A_{\rm e} = P_{\rm e} \, / \mid\!\langle\vec{S}\rangle\!\mid
 \end{equation}
the {\em effective\index{Aperture!effective} aperture} of the antenna.  $A_{\rm e}$  has the dimension of m$^2$.
Compared  to the {\em geometric\index{Aperture!geometric} aperture}
$A_{\rm g}$  an 
 aperture\index{Aperture!efficiency} efficiency $\eta_{\rm A}$  can be defined by:
 \begin{equation}
   \label{aper-eff}
   \fboxsep2mm\fbox{$ \displaystyle
       A_{\rm e} = \eta_{\rm A} A_{\rm g} $} \quad .
 \end{equation}

If an antenna with a normalized power
pattern $P_{\rm n} (\vartheta, \varphi)$ is used to receive radiation from  a
brightness distribution $B_\nu (\vartheta, \varphi)$ in the sky, at the output terminals of the antenna  the power  per unit bandwidth (PSD), in Watts Hz$^{-1}$,   $P_\nu $ is:  
 \begin{equation}
   \label{tp-output}
      P_\nu = \fract{1}{2}\, A_{\rm e} \int\!\!\int
        B_{\nu} (\vartheta, \varphi) \,P_{\rm n} (\vartheta, \varphi)
            \,\diff \Omega \; .
 \end{equation}
By definition, this operates  in the Rayleigh-Jeans limit, so the equivalent distribution of brightness
temperature can be replaced by  an
equivalent {\em antenna temperature}  $T_{\rm A}$  (Eq.~\ref{Nyquist}):  
 \begin{equation}
   \label{ant-temp}
     P_\nu = k \,T_{\rm A} \,.
 \end{equation}
This definition of {\em antenna\index{Temperature!antenna}
temperature}\index{Brightness temperature} relates the output of the antenna to the power
from a matched resistor. When these two power levels are equal, then the
antenna temperature is given by the temperature of the resistor.
The effective aperture $A_{\rm e}$ can be replaced by the 
the beam solid angle $\Omega_{\rm A} \cdot  \lambda^2$. Then (Eq.~\ref{tp-output}) becomes
 \begin{equation}
  \label{a-temperature}
  \fboxsep2.5mm
  \fbox{$\displaystyle
    T_{\rm A}(\vartheta_0, \varphi_0) =  \frac{\int
     T_{\rm B}(\vartheta, \varphi) P_{\rm n}(\vartheta - \vartheta_0,
        \varphi - \varphi_0) \sin \vartheta \diff \vartheta \diff
         \varphi}{\int P_{\rm n}(\vartheta, \varphi) \diff \Omega}  $ }
 \end{equation}
From (Eq.~\ref{a-temperature}), $T_{\rm A} < T_{\rm B}$ in all cases. The numerator   is the {\em convolution} of the brightness temperature
with the beam pattern of the telescope (Fourier methods are of great value in this analysis; see Bracewell 1986).
The brightness temperature\index{Temperature!brightness} 
 $T_{\rm b}(\vartheta, \varphi)$
 corresponds to the thermodynamic temperature of the
radiating material {\em only} for thermal radiation in the Rayleigh-Jeans
limit from an optically thick source; in all other cases $T_{\rm B}$ is
 a convenient quantity that represents source intensity at a given  frequency. 
 The quantity $T_{\rm A}$ in (Eq.~\ref{a-temperature}) was obtained for an
antenna
in which  ohmic losses and absorption in the
earth's atmosphere were neglected. These losses can be corrected in the  calibration process. 
Since $T_{\rm A}$ is the quantity measured while $T_{\rm B}$ is 
desired, (Eq.~\ref{a-temperature}) must be inverted.
(Eq.~\ref{a-temperature}) can be solved only 
 if 
 $T_{\rm A}(\vartheta, \varphi)$ and $P_{\rm n}(\vartheta, \varphi)$
are known exactly over the full range of angles. In
practice this inversion is possible only
approximately, since both  $T_{\rm A}(\vartheta, \varphi)$
and $P_{\rm n}(\vartheta, \varphi)$ are
known only for a limited range of $\vartheta$ and $\varphi$
 values, and the measured data are affected by noise.
Therefore  only an
approximate deconvolution can be  performed. If 
the source distribution
 $T_{\rm B} (\vartheta, \varphi)$
has a small extent compared to the telescope beam,  the best estimate for the upper limit to the
actual FWHP source size is 1/2 of the FWHP of the telescope beam.

\subsubsection{ Efficiencies for Compact Sources}\label{gaussians2}

 For a source small compared to the beam (Eq.~\ref{tp-output}) and (Eq.~\ref{ant-temp}) give:

 \begin{equation}
 \label{t-a-0}
   P_\nu \, = \fract{1}{2} A_{\rm e} \,S_\nu  =
        k \,T_{\rm A} 
 \end{equation}
$T_{\rm A}$ is the antenna temperature at the receiver, while  $T_{\rm A}'$  is this quantity corrected for effect of the earth's atmosphere. In the meter and cm range  $T_{\rm A}  =  T_{\rm A}'$, so in the following, $T_{\rm A}'$ will be used:
  \begin{equation}
  \label{t-a}
  \fboxsep2mm
  \fbox{$\displaystyle
           T_{\rm A}' = \Gamma S_\nu
        $}
  \end{equation}
where $\Gamma$ is the {\em sensitivity\index{Sensitivity!telescope}} of the
telescope measured in
K Jy$^{-1}$. Introducing the aperture
efficiency $\eta_{\rm A}$ according to (Eq.~\ref{aper-eff}) we find
\begin{equation}
  \label{sensit}
  \fboxsep2mm
  \fbox{$\displaystyle
         \Gamma  = \eta_{\rm A} \frac{\pi D^2}{8 k}
            $} \quad .
  \end{equation}
Thus $\Gamma$ or $\eta_{\rm A}$ can be measured with the
help of a calibrating source provided that the
diameter $D$ and the noise power scale in the receiving system
are known.
When (Eq.~\ref{t-a}) is solved for $S_\nu$, the result is: 

\begin{equation}
 \label{Tcal}
	S_\nu = 3520 \,\frac{T_{\rm A}' [{\rm K}]}{\eta_{\rm A} [{\rm D/m}]^2}
\,.
\end{equation}
  The {\it brightness\index{Temperature!brightness} temperature} is defined
as the Rayleigh-Jeans
     temperature of an equivalent black body which will give the same
     power per unit area per unit frequency interval per unit solid angle
     as the celestial source. Both $T_{\rm A}'$ and T$_{\rm MB}$ are defined in the
     Rayleigh-Jeans limit, but the brightness temperature scale has to be corrected for
     antenna\index{Temperature!antenna}
      efficiency.  
The conversion from source flux density
     to source brightness temperature for sources with sizes small
     compared to the telescope beam is given by (Eq.\,\ref{flux-den-tb2}). 
    
     For sources small compared to
     the beam, the antenna\index{Temperature!antenna}
      and main beam
     brightness temperatures\index{Temperature!brightness}
     are
     related by the main beam efficiency, $\eta_{\rm B}$:
     \begin{equation}
     \eta_{\rm B} =  \frac{T_{\rm A}'}{T_{\rm MB}} \, .
     \end{equation}
  This is valid for  sources where  sidelobe structure is not important (see the discussion after (Eq.~\ref{a-temperature})). 
   Although a source may not have a  Gaussian shape,    fits  of multiple Gaussians can be used to obtain an accurate representation.


What remains is a calibration of the temperature scales and a correction for absorption in the earth's atmosphere. This is dealt with in Section \ref{sd-cals}

 \subsubsection{Foci, Blockage and Surface Accuracy}
\begin{figure}[t]
  \includegraphics[width=11.7cm,height=4.6cm]{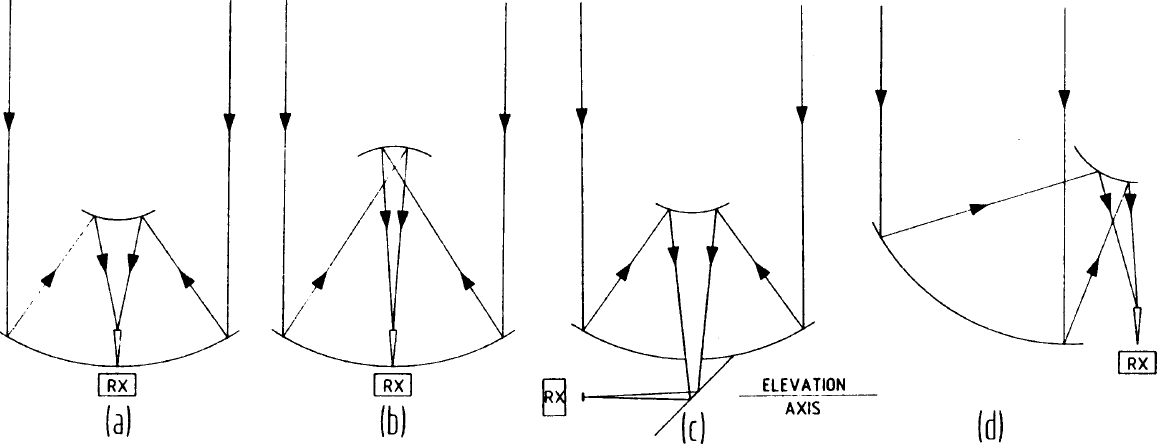}
  \caption[Geometry of Cassegrain, Gregory, Nasmyth and offset Cassegrain systems]
	{ \label{geo-cg}
   The geometry of parabolic apertures:  ({\bf a}) Cassegrain, ({\bf b}) Gregorian, ({\bf c})
   Nasmyth and
({\bf d}) offset Cassegrain systems (from Wilson et al.~2008). 
	}
\end{figure}

If the size of a radio telescope is more than a few hundred
wavelengths,
 designs are similar to those of optical telescopes. 
Cassegrain, Gregorian and Nasmyth
systems have been used.  See Fig.~\ref{geo-cg} for a sketch of these focal systems. In a
Cassegrain\index{Antenna!cassegrain } system, a convex
hyperbolic reflector is introduced into the converging beam
immediately in front of the prime focus.
This reflector transfers the converging rays to a
secondary focus which, in most
practical systems is situated close to the apex of the main 
dish. A Gregorian\index{Gregory system} system
makes use of a concave reflector with an elliptical
profile. This must be
positioned behind the prime focus in the diverging beam.
In the Nasmyth\index{Antenna!nasmyth } system this secondary focus is situated in the
elevation axis of the telescope by introducing another, usually flat,
mirror. The advantage of a Nasmyth system is that the receiver front
ends remain horizontal while  when the telescope is pointed toward
different elevations. This is an advantage for receivers cooled with
liquid helium, which may become unstable when tipped. Cassegrain and Nasmyth foci are commonly used in the mm/sub-mm wavelength ranges.

In a secondary 
reflector  system,  feed illumination  
beyond the edge   receives radiation
from the sky, which has a temperature of only a few K. For low-noise systems,
this results in only a small overall system noise temperature. This is 
significantly less than for prime focus systems. This is quantified in the so-called $''$G/T value$''$\index{Antenna!G/T value}, that is, the ratio of  antenna  gain of  to 
system noise. Any telescope design must aim to minimize the excess noise at the receiver input while maximizing gain. 
For a specific antenna, this maximization  involves the design of feeds and
the choice of foci. 
 %
%

The secondary reflector and its supports block the central parts
in the main dish from reflecting the incoming radiation,  causing
some significant  differences between the actual beam
pattern and  that of an unobstructed antenna. Modern designs seek to minimize blockage due to the support legs and subreflector.

 The  beam pattern differs  from
a uniformly illuminated unblocked aperture for 3 reasons:  

\noindent
(1) the illumination of
the reflector will not be uniform but has a taper by 10\,dB, that is, a factor of 10 or 
more at the edge of the reflector.  This is in contrast to optical telescopes which have no taper. 

\noindent
(2) the
side-lobe\index{Antenna!sidelobes} level is strongly influenced by this taper: a larger taper lowers the sidelobe level.

\noindent 
(3)  the secondary
reflector must be supported by three or four support legs\index{Antenna!feed supports}, which will
produce aperture blocking and thus  affect the shape of the beam
pattern. 

Feed leg blockage will cause deviations from circular
symmetry. For altitude-azimuth telescopes these sidelobes will
change position on the sky  with hour angle (see Reich et al.~1978). This may be a serious
defect, since these effects will be significant for maps of low
intensity regions  near an intense source.
The sidelobe response may  depend on the polarization
of the incoming radiation (see Section \ref{lr-observ}).

   A disadvantage of on-axis systems, regardless of focus, is that they
are often more susceptible to instrumental frequency baselines, so-called
{\em baseline\index{Antenna!baseline ripple} ripples} across the receiver band
than primary focus systems (see Morris 1978). Part of
this ripple is caused by reflections of noise from source or receiver 
in the antenna structure. Ripples from the receiver  can be removed if the amplitude and phase are constant in time.  Baseline ripples caused by the source, sky or ground radiation are more difficult to eliminate since these will change over short 
times. It is known that 
large amounts of blockage and  larger feed sizes lead to large baseline ripples.  
 The influence of baseline ripples on measurements  can be reduced  to a limited
extent by appropriate observing procedures.
A possible solution is an 
off-axis\index{Antenna!off-axis} system such as the GBT of the  National Radio Astronomy Observatory. In contrast to the GBT, the Effelsberg 100-m has a large amount of blocking from massive feed support legs and, as a result, show large instrumental frequency baseline ripples. These ripples 
might be mitigated  by the use of scattering cones in the reflector.

The gain of a filled aperture antenna
with small scale  surface  irregularities $\epsilon$ cannot increase indefinitely with increasing frequency but
reaches a maximum at $ \lambda_{\rm m} = 4\pi \epsilon$,
and this gain is a factor of 2.7  below that of an error-free antenna of identical dimensions.
The usual rule-of-thumb is that the irregularities should be 1/16 of the shortest wavelength used. Larger filled aperture radio telescopes are made up of panels. For these, the irregularities are of two types: (1)  roughness of the individual panels, and (2) misadjustment of panels. The second irregularity gives rise to\index{Antenna!error beam} an {\it error beam}. The FWHP of the error beam is given approximately by the ratio of wavelength to panel size. In addition, if the surface material is not a 
perfect conductor, there will be some loss and consequently  additional noise.
\subsection{Single  Dish Observational Techniques}
\label{sd-cals}
 
\subsubsection{The Earth's Atmosphere}
 \label{observations}
For ground--based facilities, the amplitudes of astronomical signals  have  been attenuated and the phases have been altered by the
earth's atmosphere. In addition to attenuation, the receiver noise is increased by atmospheric emission, the
signal is refracted and there are
changes in the path length. These effects may change slowly with
time, but there can also be rapid changes such as scintillation and anomalous
refraction. Thus propagation properties  must 
be taken into account if the astronomical measurements are to be correctly interpreted. At meter wavelengths, these effects are caused by  the ionosphere. 
 In the  mm/sub-mm 
range, tropospheric effects\index{Absorption!tropospheric} are  especially important. The
various constituents of the atmosphere absorb by different amounts.
Because the atmosphere can be considered to be in  LTE, these
constituents also  emit radiation.

The total amount of precipitable water\index{Precipitable water} (usually
measured in mm)  is an integral along the
line-of-sight to a source. Frequently, the amount of H$_2$O is determined by 
measurements of the continuum emission of the atmosphere with  a small dish at 225\,GHz. For a set of measurements at  elevations of 20$^{\rm o}$, 30$^{\rm o}$, 60$^{\rm o}$ and 90$^{\rm o}$, combined with models, rather accurate values of the atmospheric $\tau$ can be obtained. For extremely dry mm/sub-mm sites, measurements of the 183 GHz 
spectral line of water vapor  
can be used to estimate the total 
amount of H$_2$O in the atmosphere. For sea level sites, the 22.235 GHz line of water vapor has been used for this purpose.
The scale height $H_{\rm H_2 O} \approx 2\, {\rm km}$, is considerably less
than $H_{\rm air} \approx 8\, {\rm km}$ of dry air.
For this reason, sites for submillimeter radio telescopes
are usually mountain sites with elevations above $\approx 3000$\,m. For ionospheric effects, even the highest sites on earth provide no improvement.

The effect on the intensity of a radio source
due to propagation\index{Propagation effects} through  the atmosphere is
given by the standard relation for radiative transfer  (from (Eq.\,\ref{tb-isoth})): 
 \begin{equation}
\label{atm-abs}
   \fboxsep2mm
   \fbox{$ \displaystyle
     T_{\rm B}(s) = T_{\rm B}(0) \,{\rm e}^{- \tau_\nu(s)} +
          T_{\rm atm}  \, (1- \,{\rm e}^{- \tau_\nu(s)})   $} \quad .
  \end{equation}
Here $s$ is the (geometric) path length along the line-of-sight with
$s = 0$ at the upper edge of the atmosphere and $s = s_0$ at the antenna, $\tau_\nu(s)$ is the optical depth, $T_{\rm atm}$ is the temperature of the atmosphere  and $T_{\rm B}(0)$ is the temperature of the astronomical source above the atmosphere. Both the (volume) absorption\index{Volume absorption coefficient}
coefficient $\kappa$ and the
gas temperature $T_{\rm atm} $ will vary with $s$.  Introducing the mass absorption
coefficient $k_\nu$ by
\begin{equation}
\label{k}
	\kappa_\nu = k_\nu \cdot \varrho  \,,
\end{equation}
where $\varrho$ is the gas density; this variation of $\kappa$ can
mainly be related  to that of $\varrho$ as long as the gas mixture remains
constant along the line-of-sight. This is a simplified relation. For a more detailed calculations,  a multi-layer model is needed. 
 
Models can provide corrections for average effects; fluctuations and detailed corrections needed for astronomy must be determined from real-time  measurements. 

 
\vglue-10pt
\vglue-10pt

\subsubsection{Meter and Centimeter Calibration Procedures}
\label{m-cm-cal-proc}
This involves a three step  procedure: (1)  the measurements must be corrected for 
atmospheric effects, (2)  relative  calibrations are  made using secondary standards and (3) if needed, gain versus elevation curves for the antenna must be established.

In the  cm wavelength range, atmospheric effects are usually small. For steps (2) and (3) the calibration is carried out with the use of a pulsed signal injected before the receiver. This pulsed signal  is added to the receiver
input. The calibration signal must be stable, broadband and of reasonable size. Often noise diodes are used as pulsed broadband calibration sources. These are secondary standards that provide broadband radiation with effective temperatures $>10^5$~K.  With a pulsed calibration, the receiver outputs are recorded separately as: (1) receiver only, (2) receiver plus calibration and (3) repeat of  this cycle. If the calibration signal has a known value and the zero point of the receiver system is measured, the receiver noise is determined (see Eq.~\ref{tsrc}).  Most often the calibration value in either Jy/beam  or T$_{\rm MB}$ units is determined by a continuum scan through a non-time variable compact discrete source of known flux density. Lists of calibration sources are to be found in 
 Baars et al.~(1977), Altenhoff (1985), Ott  et~al. (1994) 
and  Sandell \,(1994).  

\subsubsection{Millimeter and Sub-mm Calibration Procedures}\label{mm-cal-proc}

In the mm/sub-mm wavelength range, the atmosphere has a
larger influence and can change on timescales of seconds, so more complex 
corrections are needed. Also,large telescopes may operate close to the limits caused by their
surface accuracy, so that the power received in the error beam may be
comparable to that received in the main beam. In addition, many sources such as molecular clouds are rather extended. Thus, relevant values of 
telescope efficiencies must be used (see Downes 1989).  The calibration procedure used in the mm/sub-mm range is
referred to as the {\em chopper wheel} method (Penzias \& Burrus 1973). This  consists of two steps:

\noindent 
(1) the measurement of the receiver noise (the method is very similar to that in  Section (\ref{cal-proc}). 
 and 

\noindent 
(2) the measurement of the
receiver response when directed toward cold sky at a
certain elevation. 

In the following it is assumed that the receiver is operated in the SSB mode.  For (1), the output of the receiver while measuring
an ambient load, $T_{\rm amb}$,  is denoted by $V_{\rm amb}$:

\begin{equation}
	  \label{vamb}
	V_{\rm amb}= G\,(T_{\rm amb}+T_{\rm rx}) \, .
\end{equation}
where $G$ is the system gain. This is sometimes repeated with a second load at a different temperature. The result is a determination of the receiver noise as in  Section (\ref{cal-proc}). 
For step (2), the load is removed; then the output refers to noise from a source-free sky ($T_{\rm sky}$), ground ( $T_{\rm gr} = T_{\rm amb}$) and receiver: 
\begin{equation}
  \label{vsky}
	  V_{\rm sky} =
	G\,[F_{\rm eff} \,T_{\rm sky} +  (1 - F_{\rm eff}) \,T_{\rm gr} +
T_{\rm rx}] \,.
\end{equation}
where $F_{\rm eff}$ is  the {\em forward efficiency}. This is
 the fraction of power in the forward beam of the feed. This can be interpreted as the response to a source with the angular size of the Moon (it is assumed that $F_{\rm eff}$  is appropriate for an extended molecular cloud).  
Taking the difference between $V_{\rm amb}$ and $V_{\rm sky}$: 
\begin{equation}
  \label{vcal}
\Delta  V_{\rm cal} = V_{\rm amb}-V_{\rm sky}=G \,F_{\rm eff} \,T_{\rm amb}
                \expo^{-\tau_\nu} \, ,
\end{equation}
where $\tau_\nu$ is the atmospheric absorption at the frequency of
interest. If it is assumed that  $ T_{\rm sky}(s) =           T_{\rm atm}  \, (1- \,{\rm e}^{- \tau_\nu})   $ describes the emission of the atmosphere, and, as in (Eq.~\ref{atm-abs}),   $\tau_\nu$ in  is the same for emission and absorption, emission measurements can provide the value of $\tau_\nu$. If $T_{\rm atm}= T_{\rm amb}$, the correction is simplified.  For more complex situations, models of the atmosphere are needed 
(see e.g., Pardo et al.~2009). Once $\tau_\nu$ is known, the  signal  from the radio
source, $T_{\rm A}$, after passing through the earth's atmosphere, is
\beq
	\Delta V_{\rm sig}=G \,T_{\rm A}' \,  {\rm  e}^{-\tau_\nu}
\eeq
or
\beq
	T_{\rm A}'=\frac{\Delta V_{\rm sig}}{\Delta V_{\rm cal}} \,F_{\rm eff}
             \,T_{\rm amb}
\eeq
where $T_{\rm A}'$ is the antenna
temperature\index{Temperature!antenna}\index{Temperature!brightness}
 of the source outside the
earth's atmosphere. We define
\begin{equation}
 \label{corta}
	T_{\rm A}^*=\frac{T_{\rm A}'}{F_{\rm eff}} =
	\frac{\Delta V_{\rm sig}}{\Delta V_{\rm cal}} \,T_{\rm amb} \, 
\end{equation}
The right side involves only measured quantities.  $T_{\rm A}^*$ is commonly referred to as the {\em corrected
antenna temperature},\index{Temperature!antenna}
 but it is really a {\em forward beam brightness
temperature}.\index{Temperature!brightness}
 An analogous temperature is $T_{\rm sys}^*$, the system noise correcting for all atmospheric effects: 

\begin{equation}
  \label{T-sys-star}
	T_{\rm sys}^*=\left(\frac{T_{\rm rx} + T_{\rm sky}}{F_{\rm eff}} \right) \, {\rm e}^{\tau}
\end{equation}
This result is used to determine continuum or  line temperature scales (Eq.~\ref{correl-cal-base}). A typical set of values  for $\lambda=3$mm are:  $T_{\rm rx} $=40~K, $T_{\rm sky}$=50~K, $\tau$=0.3. Using these, the  $T_{\rm sys}^*$=135~K. 

For sources $\ll $30\arcmin, 
there is an additional correction  for the telescope beam efficiency, which is commonly
referred to as $B_{\rm eff}$.
Then
\beq
	T_{\rm MB}=\frac{F_{\rm eff}}{B_{\rm eff}} \,T_{\rm A}^*
\eeq
Typical values of $F_{\rm eff}$ are $\cong 0.9$, and at the shortest wavelengths used for a telescope, $B_{\rm eff} \cong 0.6$. In general, for extended sources,  the brightness temperature corrected for absorption by the earth's atmosphere, 
$T_{\rm A}^*$,  should be used. 
 
\subsubsection{Bolometer Calibrations}\label{incoherent-cals}

Since most bolometers are A.~C. coupled (i.~e.~respond to differences), so the D.~C. response (i.~e.~respond to total power)  used in   $''$hot--cold$''$ or $''$chopper wheel$''$ 
calibration methods cannot be used. Instead astronomical data are calibrated in two steps: 

\noindent 
(1) measurements of atmospheric emission at a number of elevations to 
determine the opacities at the azimuth of the target source, and

\noindent 
(2)  the measurement of  the response of a nearby source with a known flux density; immediately  after this, a  measurement of the target source is carried out.

\subsubsection{Continuum Observing Strategies}\label{cont-obs}
 
\vglue-3pt
 
\noindent
{\bf 1) Position Switching and Wobbler Switching.}  Switching against a load or absorber  is  used only in exceptional circumstances, such as  studies of the 2.73 ~K cosmic microwave background\index{Cosmic Microwave Background}. For the CMB, Penzias \& Wilson (1965) used a helium cooled load with a precisely known temperature. 
  For compact regions, compensation of transmission variations of the atmosphere
is possible if double \index{Double beam system} beam systems can be used. At higher frequencies, in the mm/sub-mm range, rapid movement  
of the telescope beam (by small movements of the sub-reflector or a mirror in the path from  antenna to receiver) over small angles is referred to as  $''$beam switching$''$,   $''$wobbling\index{Wobbling
scheme}$''$ or $''$wobbler switching$''$. This   is used to produce two beams on the sky for a single pixel receiver.  
The
individual telescope beams should be spaced by a distance of 
3~FWHP beam widths. 

\noindent 
{\bf 2) Mapping of Extended Regions and  On the Fly Mapping.} 
Multi-beam bolometer systems are preferred for continuum measurements at $\nu > $ 100 GHz.  Usually, a wobbler system is
needed for such arrays.  
With these, it is possible to measure a fairly large  region and  to  better cancel  
sky noise due to weather. Some details of more recent  data taking and reduction methods are given in e.g.,~Johnstone et al.~(2000) or Motte et al.~(2006).

If extended areas are to be mapped, scans are made along one direction (e.g., Azimuth or Right Ascension). Then the antenna is offset in the orthogonal direction by 1/2 to 1/3 of a  beamwidth, and the scanning is repeated until the region is 
completely mapped. This is referred to as a $''$raster
scan$''$\index{Raster scan}. There should  be reference positions free of sources at the beginning and
the end of each scan, to allow the determination of zero levels and  calibrations should be made before the scans are begun. For more secure results, the map is then repeated by scanning in the orthogonal direction (e.g., Elevation or Declination). Then both sets of results are placed on a common   grid, and averaged; this is referred to as  $''$basket\index{Basket weaving} weaving$''$.

Extended emission\index{Emission!extended regions} regions can also
be mapped using a double beam system, with the receiver input periodically switched between the first and second beam. In this procedure,  there is some suppression
of very extended emission. A  summation of the beam switched data along the scan direction has been
used to reconstruct infrared images. More sophisticated schemes  can recover most, but not all, of the
information (Emerson et al. 1979; $''$EKH$''$). Most mm/sub-mm antennas employ  wobbler switching in azimuth to
cancel ground radiation. By measuring a source using scans in azimuth at
different hour angles,  then transforming the positions to an astronomical coordinate frame and combining the maps 
it is possible to reduce the effect  of sidelobes caused by feed legs and supress sky noise (Johnstone et al.~2000). 
 
\subsubsection[Additional Requirements for Spectral Line Observations]{Additional 
Requirements\protect\newline for Spectral Line Observations}
 \label{lr-observ}
 
In addition to the requirements placed on continuum receivers, there
are three  additional requirements for spectral line receiver systems.

If the observed frequency of a line is
compared to the known rest frequency, the
relative radial\index{Velocity!radial} velocity of the  source and the
receiving system can be determined. But
this velocity contains the motion of the
source as well as that of the receiving
system, so the velocity measurements are referred to some
standard of rest. This velocity can be separated into several
independent components:  (1) Earth rotation with a maximum  velocity $v = 0.46 $ km\,s$^{-1}$ and  (2) The motion of the center of the Earth  
relative to the barycenter of the 
Solar System is  said to be reduced to the {\em heliocentric\index{Radial velocity!heliocentric}} system. Correction  algorithms are available 
for  observations of the earth relative to center of mass of the
solar system. 
The  {\em standard
solar\index{Solar motion!standard} motion}  is the 
motion relative to the mode of the velocity of the stars in
the solar neighborhood. Data where the standard
solar motion has been taken into account are said
to refer to the {\em local standard\index{Local standard of rest} of
rest} (LSR\index{LSR}). Most extragalactic spectral line  data do {\em  not} include the LSR correction but are referred to the heliocentric velocity.  For high redshift sources, special relativity corrections must be included.

For larger bandwidths, there is an instrumental spectrum and  a
$''$baseline\index{Baseline}$''$
must be subtracted from the (on-off)/off  spectrum. Often a linear
fit to spectrum  is sufficient,
but if curvature \index{Baseline!curvature} is present,  polynomials of second or
higher order must be subtracted.
At high galactic latitudes, more intense 21 cm line radiation from the galactic plane can give rise to artifacts in spectra from scattering of radiation within the antenna (see Kalberla et al. 2010). This is apparently less of a problem in surveys of galactic carbon monoxide (see Dame et al.~1987).

\subsubsection{ Spectral Line Observing Strategies}
 \label{splotech}
 
\vglue-2pt\noindent
Astronomical  radiation is 
often only a small fraction of the total
power received. 
To avoid  stability problems,  the
signal of interest must be compared with another 
that contains approximately the same total power and
differs only 
that it contains no source. The receiver must be stable so that any gain or 
bandpass changes occur over time scales
 long compared to the time needed
for position change. To
detect an astronomical source, three  observing modes 
are used to produce a 
suitable comparison.
 
 


\vspace*{3ex}
\noindent
{\bf 1) Position Switching and Wobbler Switching.}  \index{Position switching} The signal
$''$on source$''$ is compared with a measurement 
obtained at a nearby position in the sky.
 For spectral lines, there must be 
little line radiation at the comparison
region.  This is referred to as the $''$total power$''$ observing mode\index{Total power observing}.
A variant of this method is wobbler switching. This is very useful for compact
sources, especially in the mm/sub-mm range.
\enlargethispage{\baselineskip}  
 
\vspace*{3ex}
\noindent
{\bf 2) On the Fly Mapping.} This  very important observing
method is an extension of method (1)\index{On-the-fly mapping}. 
In this procedure, 
 spectral line data is taken at a rate of perhaps one spectrum or more per
second.

\vspace*{3ex}
\noindent
{\bf 3) Frequency Switching}\index{Frequency switching}.
For many sources, spectral line radiation  at  $\nu_0$  is restricted to a narrow band, that is,  present only over a small frequency 
interval, $\Delta \nu$, for example  $\Delta \nu  / \nu_0  \approx 10^{-5}$. If  all other effects vary very
little over $\Delta \nu $,  changing  the frequency of a 
receiver on a short time by up to  $10 \, \Delta \nu $  produces
a comparison signal with the  line well shifted. The line is measured all of the time, so this is an efficient observing mode. 
%



\section{Interferometers and Aperture Synthesis  }
     \label{interferometers}
 
From  diffraction theory,  the angular
\index{Resolution!angular} resolution 
 is  given by (Eq.~\ref{diffraction}). 
 However, as shown by Michelson (see Jenkins \& White 2001),   
a much higher  resolving power  can be obtained
by coherently combining the output of two  reflectors of diameter $d\ll B$ separated by a distance $B$ yeilding  $\theta \approx \lambda$/B. 
In the radio/mm/sub-mm range, from (Eq.~\ref{heisenberg6}),  the outputs can be amplified without seriously degrading the signal-to-noise ratio. This amplified signal  can be  divided and used to produce a large number of cross-correlations.


Aperture synthesis is a further development. This is the procedure to produce  high quality images of  sources  by combining a
 number of measurements  for different  antenna  spacings up to  the maximum $B$. The longest spacing  gives the angular resolution of 
an equivalent  large  aperture. This has become  the  method  to obtain high
quality, high angular resolution images. The first practical demonstration of aperture synthesis in radio astronomy was made by M. Ryle and his associates (see Section 3 in Kellermann \& Moran 2001). 
Aperture synthesis allows the  reproduction of the imaging properties of a large aperture by sampling the radiation field at
individual positions within the aperture.  
Using this approach, 
 a remarkable improvement of the radio astronomical imaging  was made possible. 
More detailed 
accounts are to be found in Taylor et al.~(1999),  Thompson et al.~(2001)  or Dutrey (2001).


The simplest case is  a two element system in which electromagnetic waves are received by two antennas. These induce the voltage $V_1$ at $A_1$:
 \begin{equation}
   \label{out-ua}
    V_1 \propto E \expo^{\,\im \omega t} \,,
 \end{equation}
while at  $A_2$:
 \begin{equation}
   \label{out-ub}
    V_2 \propto E \expo^{\,\im \omega \,(t - \tau)} \,,
 \end{equation}
where $E$ is the amplitude of the incoming electromagnetic plane wave, $\tau$ is the geometric delay caused by the relative orientation of the
interferometer baseline $\vec{B}$ and the direction of the wave propagation.
For simplicity, 
receiver noise and instrumental phase were neglected in (Eq.~\ref{out-ua}) and (Eq.~\ref{out-ub}). The outputs will be
correlated. Today all radio interferometers  use direct  correlation  followed by an integrator. 
\begin{figure}
  \includegraphics[width=5.7cm,height=5.8cm]{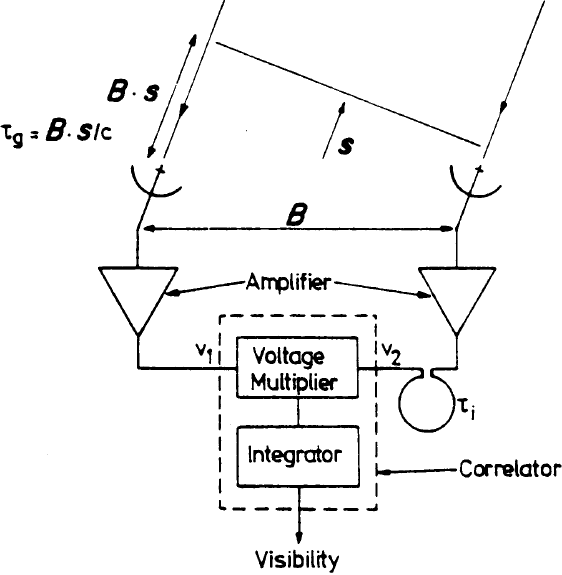}      
  \caption[Two element correlation interferometer]
	{ \label{two-el-cor}
	A schematic diagram of a two element correlation interferometer.
	The antenna output voltages are $V_1$ and $V_2$; the instrumental 
	delay is $\tau_{\rm i}$ and the geometric delay is $\tau_{\rm g}$.  $\vec{s}$ is the direction to the source. Perpendicular to $\vec{s}$ is the projection of the baseline $\vec{B}$. The signal is 
digitized after conversion to an intermediate frequency. Time delays are introduced  using digital shift registers  (from Wilson et al.~2008). 
	}
\end{figure}
%
The output is proportional to: 
\beq
   R (\tau) \propto \frac{E^2}{T} \int\limits_0^T \expo^{\,\im \omega t}
                \expo^{\,-\im \omega ( t - \tau)} \,\diff t \, .
\eeq
If $T$ is a time much longer than the time of a single full
oscillation, i.e., $ T \gg 2 \pi / \omega$
then the average over time $T$ will not differ much
from the average over a single full period, 
resulting in
\begin{equation}
   \label{r-tau}
  \fboxsep2mm\fbox{$ \displaystyle
    R (\tau) \propto \fract{1}{2} E^2 \expo^{\,\im \omega \tau} $}
\quad .
 \end{equation}

The output of the correlator\,$+$\,integrator  varies
periodically with $\tau$, the delay. 
Since $\vec{s}$ is slowly changing due to the rotation of the
earth,  $\tau$ will vary, producing 
{\em interference fringes} as a function of time.
 
The basic components of a
 two element system are
shown in Fig.\,\ref{two-el-cor}.
 If the radio brightness distribution is given by $I_\nu(\vec{s})$, the
power received per bandwidth $\diff \nu$ from the source element
$\diff \Omega$ is $ A(\vec{s}) I_\nu(\vec{s}) \diff \Omega \diff \nu$,
where $A(\vec{s})$ is the effective collecting area in the direction $\vec{s}$;
 the same $A(\vec{s})$ is assumed  for each of the antennas.
The amplifiers are assumed to have   constant gain and phase factors (neglected here for simplicity).
 
The output of the correlator for radiation from the direction $\vec{s}$
(Fig.\,\ref{two-el-cor})  is
 \begin{equation}
   \label{rot}
    r_{12} = A(\vec{s}) \,I_\nu(\vec{s}) \,\expo^{\im \omega \tau}
               \,\diff \Omega \diff \nu
 \end{equation}
where $\tau$ is the difference between the geometrical and instrumental
delays $\tau_{\rm g}$ and $\tau_{\rm i}$.
If $\vec{B}$ is the baseline vector between  the two antennas
 \begin{equation}
    \label{delay}
    \tau = \tau_{\rm g} - \tau_{\rm i} = \frac{1}{c} \,\vec{B} \cdot \vec{s}
              - \tau_{\rm i}
 \end{equation}
 the total response is obtained by integrating over the source $S$
 \begin{equation}
   \label{tot-res}
   \fboxsep3mm\fbox{$ \displaystyle
     R(\vec{B}) =  \parbox{8mm}{$ \vspace*{-2mm}
                       {\displaystyle
                        \int\!\!\int \atop
                         {\!\!\!\!\!\scriptstyle \Omega }} $ }
       A(\vec{s}) I_\nu(\vec{s}) \expo^{ 2 \pi  \im \nu \left(\frac{1}{c}
            \,\vec{B} \cdot \vec{s} - \tau_{\rm i} \right) }
             \diff \Omega \diff \nu  $}    \quad 
 \end{equation}
The function $R(\vec{B})$, the {\em Visibility Function}  is closely related to
the mutual coherence function (see  Born \& Wolf 1965, Thompson et al. 2001, Wilson et al. 2008) of the source. For parabolic antennas,  it is usually assumed that $A(\vec{s}) = 0$
outside the main beam area so that  (Eq.~\ref{tot-res})
is integrated only over this region.
A one dimensional version of (Eq.~\ref{tot-res}), for  a baseline $B$, frequency $\nu=\nu_0$ and instrumental time delay  $\tau_i=0$, is
 \begin{equation}
   \label{1d-res-simple}
     R(B) =    \int   A(\theta) \, I_\nu(\theta) \expo^{  2 \pi  \im \nu_0 \left(\frac{1}{c}  \,B \, \theta  \right) }
             \diff \theta 
 \end{equation}
 With  $\theta=x$ and $B_x/ \lambda=u$, this  is
 \begin{equation}
   \label{1d-res-simple1}
     R(B) =    \int   A(\theta) \, I_\nu(\theta) \expo^{  2 \pi \im  u \, x   }
             \diff \theta 
 \end{equation}
This   form of  (Eq.~\ref{tot-res})  illustrates more clearly the Fourier Transform relation of $u$ and $x$. This simplified version will be  used to provide  illustrations  of   interferometer responses (see Section \ref{int-responses}). 
In two dimensions, (Eq.~\ref{tot-res}) takes on a similar form with the additional variables $y$ and  $B_y/ \lambda=v$. The image can be obtained from the inverse Fourier transform of Visibilities;  see (Eq.~\ref{i-str}). 

 \subsection{Calibration}

Amplitude and  phase  must be calibrated for all interferometer measurements. In addition, the instrumental  passband  must be calibrated for spectral line measurements. 
The amplitude scale is calibrated by a determination of the system noise  at  each antenna using methods presented  for single dish measurements (see Section \ref{m-cm-cal-proc} and following).  In the centimeter range, the atmosphere plays a small role while in the mm and sub-mm wavelength ranges, the atmospheric 
effects must be accounted for. For phase measurements, a suitable point-like source with an accurately known position is required to 
determine  $2\pi \nu \tau_i$ in 
(Eq.~\ref{tot-res}). 
For interferometers, the best calibration sources are usually unresolved or  point-like sources. Most often these are extragalactic time variable sources.  
 To calibrate the response in units of flux density or brightness temperatures,  these amplitude  measurements must be referenced to primary calibrators (see a list of non-variable sources of known flux densities  in Ott et al.~1994 or Sandell 1994). 

The calibration of the instrumental passband is carried out by a longer integration on  an intense source to determine the channel-to-channel gains and 
offsets. The amplitude, phase and passband 
calibrations are carried out before the source measurements. The passband calibration is usually carried out every few hours or once per observing session. The 
amplitude and phase calibrations are made 
more often; the time between such calibrations  depends on the stability of the electronics and weather. If weather conditions require  frequent measurements of calibrators (perhaps less than once per minute for $''$fast switching$''$), integration time is reduced. In case of even  more rapid weather changes, the ALMA project will make use of water vapor radiometers mounted on each antenna (see Section \ref{observations}). These will be used to determine the total amount of H$_2$O vapor above each antenna, and use this to make corrections to phase.

\subsection{Responses of  Interferometers}
\label{int-responses}
\subsubsection{Time Delays and Bandwidth}
The instrumental response is reduced if 
the bandwidth at the correlator  is large compared to the delay caused by the
separation of the antennas. 
For large bandwidths, the loss of correlation  can be minimized
by adjusting the phase  delay  so that the  difference of arrival time between antennas  is negligible.
In practice, this is done by inserting a  delay between
the antennas so that $\frac{1}{c}
            \,\vec{B} \cdot \vec{s}$ equals $\tau_{\rm i}$. This
 is equivalent to centering the response on  the
central, or {\em white light fringe}.
 %
Similarly, the reduction of the response  caused by finite bandwidth can be
estimated by an integration of  (Eq.~\ref{tot-res}) over frequency, taking $A(\vec{s})$ and $I_\nu(\vec{s})$ as constants.
The result is a 
factor, $ \sin (\Delta \nu \tau)/ \Delta \nu \tau \,$ in (Eq.~\ref{tot-res}).
This will reduce the interferometer response if
$\Delta \nu \tau \sim 1$ . For
typical bandwidths of 100\,MHz, the offset from the zero delay
must be $\ll 10^{-8}$\ts s. This adjustment of delays is
referred to as  {\em fringe\index{Fringe!stopping} stopping}. The exponent in (Eq.~\ref{tot-res}) has both sine and cosine components, but  digital cross-correlators 
 record both components, so that   the entire response can be recovered. 
\subsubsection{ Beam Narrowing}

The {\em white light fringe} the delay  compensation must be set with a high accuracy to prevent a reduction in the interferometer response. 
For a finite  primary 
antenna beamwidth, $\theta_b$, this cannot be the case over the entire beam.  
For a bandwidth $\Delta \nu$  there will be a phase difference. 
 Converting the wavelengths to frequencies and  using $\sin{\theta} \cong \theta$ 
the result is
\begin{equation}
   \label{band-beam}
\Delta \phi = 2 \pi \, \frac{\theta_{\rm offset}}{\theta_b} \, \frac{ \Delta \nu}{\nu}
 \end{equation}

This effect can be  important for  continuum measurements made with large bandwidths, but  can be reduced if the cross correlation is  carried 
out using a series of narrow 
contiguous IF sections. For each of these IF sections, an extra delay is introduced to center the response at the value which is appropriate for 
that wavelength before correlation. 

\subsubsection{Source Size}
 From   an idealized  source, of shape  $I(\nu_0)=I_0$ for $\theta \, < \, \theta_0$ and  $I(\nu_0)=0$  
for $\theta \, > \, \theta_0$;  we 
take the primary beamsize of each antenna  to be much 
larger, and define the fringe width for a baseline $B$ $\theta_b$ to be $  \frac{\lambda}{B} $, 
The result is 
 \begin{equation}
   \label{1d-res-1}
 R(B) = A \, I_0 \cdot \theta_0 \, \expo^{  \im  \pi \frac{\theta_0}{\theta_b}  } \, \left[ \frac{\sin{(\pi \theta_0/\theta_b)}}{{(\pi \theta_0/\theta_b)}}  \right]
 \end{equation}
The first terms are normalization and phase factors. The important  term is  in  brackets.  
If $\theta_0  >>  \theta_b$, the interferometer 
response is reduced. This is sometimes referred to as the problem of  $''$missing short spacings$''$'. 
To correct for the loss of source flux density,  the interferometer data must be  supplemented by  single dish
measurements.  The diameter of the single dish antenna should be larger than
the shortest    interferometer spacing. This single dish image
must extend to the FWHP of the smallest of the interferometer antennas. When  Fourier transformed and appropriately combined with the interferometer response, the resulting   
data set has {\em no} missing flux density.

 \subsection{Aperture Synthesis }
 \label{apsyn}
 To produce an image, the integral equation (Eq.~\ref{tot-res})  must be inverted.  A number of approximations may have to  be 
 applied to produce high quality images. In addition,  the data are affected by noise.  The most important steps of this development will be presented.

For  imaging over a limited region of the sky rectangular coordinates are adequate, so relation (Eq.~\ref{tot-res}) can be rewritten with coordinates $(x, y)$ in the image plane  and coordinates $(u, v)$ in the Fourier plane. The coordinate $w$, corresponding to the difference in height, is set to zero. Then the relevant relation is:
 \begin{equation}
   \label{i-str}
   \fbox{$ \displaystyle
     I' (x, y) = A (x, y)\, I (x, y) = \int\limits_{-\infty}^\infty
        V ( u, v, 0) \expo^{-2 \pi \im ( u x + v y)} \diff u \diff v
        $} \quad 
 \end{equation}
where $I'(x, y)$ is
the intensity $I (x, y)$  modified by the primary beam shape $A (x, y)$. 
It is easy to correct $I' (x, y)$ by  dividing  by
$A(x, y)$.  Usually data present beyond the half power point is excluded.

 \vspace{0.5mm}

The most important definitions are:

\noindent
(1) {\em Dynamic Range}: The \index{Image!dynamic range}ratio of the maximum to the minimum intensity in 
an image. In images made with an interferometer array, it is assumed that  corrections for primary 
beam taper have been applied. If the minimum intensity is determined by the random noise in an image, the dynamic range is defined by the signal-to-noise ratio of the brightest feature in the image.
The dynamic range is an indication of the ability to recognize low intensity features in the presence of intense features.  
If the minimum noise is determined by artifacts, i.e., noise in excess of the theoretical value,  $''$image improvement techniques$''$ should be applied.

\noindent
(2) {\em Image Fidelity}: This\index{Image!fidelity} is defined by the agreement between the measured results and the actual ($''$true$''$) source structure. 
A quantitative assessment of fidelity is: 
\beq
F=|(S-R)|/S
\eeq
where $F$ is the fidelity, $R$ is the resulting image obtained from the measurement, and $S$ is the actual source structure. The highest fidelity is $F=0$. Usually errors can only be  estimated using a priori knowledge of the correct source structure.  In many  cases, $S$ is a source model, while $R$ is  obtained by  processing $S$ with a model of the instrumental response. This relation can only be applied when the value of $R$ is more than 5 times the RMS noise.

\begin{figure}
 \includegraphics[width=\linewidth]{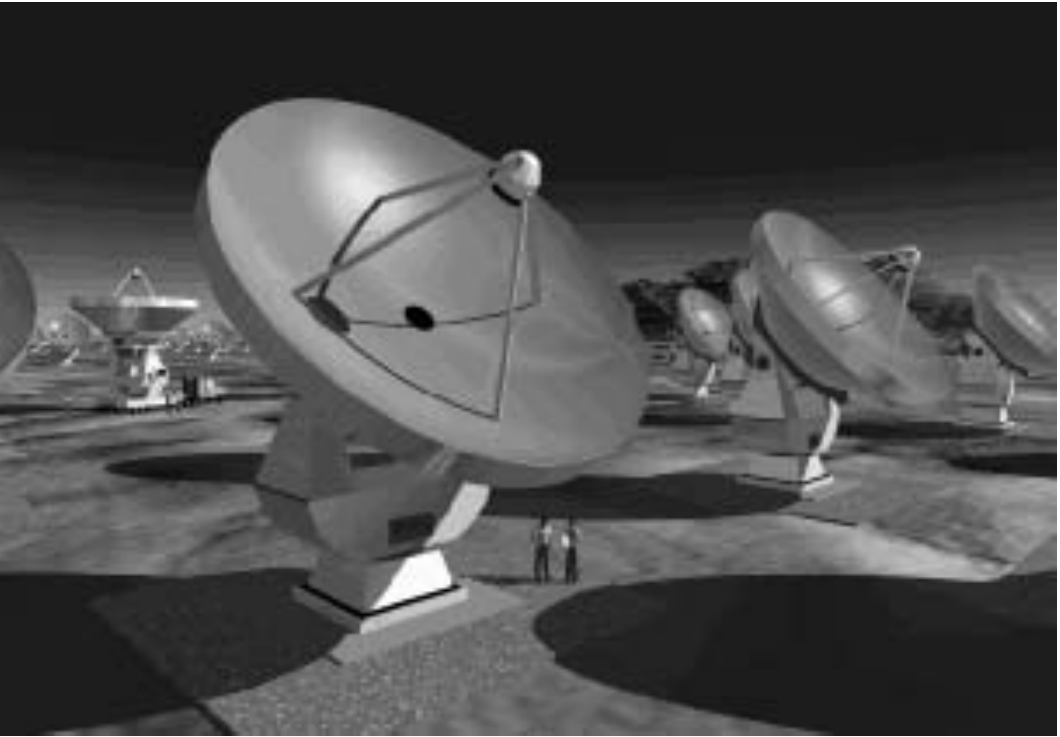}
  \caption[The ALMA Array]{ \label{ALMA}
An artists sketch of ALMA. To date, this is the most ambitious construction project in ground based astronomy. ALMA is now being 
built in north Chile on a 5 km high site. It will consist 
of fifty-four 12-m and twelve 7-m antennas, operating in 10 bands 
between wavelength 1~cm and 0.3~mm. In Early Science, four receiver bands at 3, 1.3, 0.8 and 0.6 mm will be available. The high ALMA sensitivity is due to the  
extremely low noise receivers, the  highly 
accurate antennas, and the high altitude site.  At the largest antenna spacing, 
and shortest wavelength, the angular resolution will be $\sim$5 
milliarcseconds\index{Interferometer!ALMA} (courtesy ESO/NRAO/NAOJ). }
\end{figure}

\subsubsection{Interferometric Observations}
 \label{imagerest}
 
\vglue-5pt\noindent
Usually measurements are  carried out in 1 of 4 ways.

\noindent 
1.  {\em Measurements of a single target source}. This is  similar to the case
of  single telescope position switching. Two significant differences with single dish measurements are that the interferometer 
measurement may have to extend over a wide range of hour angles 
to provide a better coverage of the $(u, v)$ or Fourier  plane,  and that instrumental phase must be determined also. 
 After the  measurement of a calibration
source\index{Calibration!source} or reference source, which has a known
position and size, the effect of instrumental phases in the
instrument and atmosphere is removed and a calibration of  the amplitudes of the source is made. Target sources and calibrators are usually observed alternately; the calibrator  should be  close to the target source. 
The time variations caused by instrumental and weather effects must be slower than the time between measurements of source and calibrator.  
If, as is the case for mm/sub-mm wavelength measurements, 
weather is an important influence,  
 target and calibration source must be measured often. For ALMA (see Fig.~\ref{ALMA}), observing will  follow a  two part scheme. 
For {\em fast switching} there will be integrations of perhaps  10 seconds on a nearby calibrator, then a few minutes on-source\index{Interferometer!fast switching}. This method will 
reduce the amount of phase fluctuations, at the cost of on-source observing time. For more
rapid changes in the earth's atmosphere, phases will be   corrected using 
measurements of atmospheric water vapor from measurements of the 183 GHz line. 

\noindent 
2.   {\em Snapshot Mode}. A  series of short observations (at different hour angles) of one source after another, and then the  measurements are repeated. 
For sensitivity reasons, snapshots  are usually made
in the radio continuum or more intense spectral lines. As in observing method (1), measurements of source and calibrator are  interspersed  
to remove the effects of instrumental phase drifts
 and to calibrate the amplitudes of the sources in
question.
The images will  affected by the shape of the synthesized
beam \index{Interferometer!synthesized beam}
since there is sparse  coverage in the $(u, v)$ plane. If the size of the source to be  imaged is comparable
to the primary beam of the individual antennas there should be a correction for the power pattern..  %

\noindent 
3.  {\em Multi-Configuration Imaging} Here the goal is the 
image of a source either with  high dynamic range or high
sensitivity. 
 Measurements with a number of different
interferometer configurations  better fill the $uv$ plane. These measurements are taken at different epochs  and after calibration, 
the visibilities are entered into a common data set.

\noindent 
4.  {\em Mosaicing}
An extension of procedure (1)  can be used  for sources with an extent much larger than the primary antenna beam. These images require  measurements at  adjacent pointings. This is spoken of as {\em mosaicing}. In a mosaic, the antennas are pointed at narby positions. These positions should overlap at the half power power point. The images can be formed separately and then combined to
produce an image of the larger region. Another method is to combine the data in the $(u, v)$ plane and then form the image.

\subsection{Interferometer Sensitivity}\label{intsens}
 
The random noise\index{Interferometer!noise} limit to an interferometer system can be 
calculated
following the method used 
for a single telescope (Eq.~\ref{dicke}). The use of  (Eq.~\ref{t-a-0}) provides a conversion from  $\Delta T_{\rm RMS}$ to $\Delta S_\nu$\index{Flux density}.  
collecting area of a single antenna. 
 For an array of $n$ identical antennas, there are
     $N= n (n-1) / 2 $ simultaneous pairwise correlations, so 
     the RMS variation in flux
     density is: 
     \begin{equation}
     \label{fluxrmssyn}
     \Delta S_\nu = \frac{2 \,M \,k \, T_{\rm sys}^*}{ A_{\rm e} \sqrt{2 \,N
\,t \,\Delta \nu}}  \,.
     \end{equation}
     with M$\cong 1$, $A_{\rm e}$  the effective area of each antenna and $T_{\rm sys}^*$ given by (Eq.~\ref{T-sys-star}). This relation can be recast in the form of brightness temperature
      fluctuations\index{Temperature!fluctuations} using the
Rayleigh-Jeans relation; then the RMS noise in  brightness temperature units is: 
     \begin{equation}
     \label{temprmssyn}
     \Delta \Tb = \frac{2 \,M \,k \,\lambda^2 \,\Tsys^*}{A_{\rm e}  \Omega_{\rm b}
     \sqrt{2 \,N \,t \,\Delta \nu}} \,.
     \end{equation}
     For a Gaussian beam, $\Omega_{\rm mb}=1.133 \, \theta^2$, so       the RMS temperature fluctuations can be  related to observed properties of a
     synthesis image.

 Aperture \index{Aperture synthesis} synthesis is based on discrete 
 samples of the visibility function $V (u, v)$, with  
the goal  of  the  densest possible  coverage  of the $(u,v)$ or Fourier  plane.
It has been observed that the RMS noise in a synthesis image obtained by
Fourier transforming the $(u,v)$ data is often  
 higher than  given by 
(Eq.~\ref{fluxrmssyn}) or  (Eq.~\ref{temprmssyn}).  Possible causes are: (1)   phase fluctuations caused by   atmospheric or instrumental instabilities,  (2) 
 incomplete 
sampling  of  the $(u,v)$ plane, which gives  rise to artifacts such as  stripe-like features in
the  images, or (3)  
 grating rings around more intense sources; these are analogous to
{\em high sidelobes\index{Antenna!sidelobes}} in single dish diffraction
patterns.

\subsection{ Corrections of  Visibility Functions}
\subsubsection{Amplitude and Phase Closure}
The relation between the
measured $\widetilde{V_{ik}}$ visibility and  {\em actual}
visibility $V_{ik}$  is  considered linear: 
 \begin{equation}
   \label{Vcal}
     \widetilde{V_{ik}}(t) = g_i(t) \,g^*_k(t) \,V_{ik} + \epsilon_{ik}(t) \; .
 \end{equation}
Values for the complex antenna gain factors $g_k$ and the noise term
$\epsilon_{ik}(t)$ are determined by measuring calibration sources as
frequently as possible. Actual values for $g_k$ are then computed
by linear interpolation. 
The (complex) gain of the
array is obtained by the multiplication of the  gains 
of the individual
antennas. If the array consists of $n$ such antennas, $n(n-1)/2$
visibilities can be measured simultaneously, but only $(n-1)$ independent
gains $g_k$ are needed  since one antenna in the array  can be taken as a
reference. So in an array with many antennas, the number of antenna pairs
greatly exceeds  the number of antennas. For phase, one must determine  $n$ phases. 
Often these conditions can be introduced into the solution in the form
of {\em closure \index{Error!closure} errors}. Defining  the phases
$\varphi, \theta$ and
$\psi$ by \\
\parbox{9cm}
{\begin{displaymath}
\begin{array}{rcl}
\widetilde{V_{ik}} &=& |\widetilde{V_{ik}}| \,\expo^{ \im \varphi_{ik} } \,, \\
G_{ik} &=& |g_i| \,|g_k| \, \expo^{  \im \theta_i }
\expo ^{- \im \theta_k } \,, \\
V_{ik} &=& |V_{ik}| \,\expo^{ \im \psi_{ik} } \,. \\
\end{array}
\end{displaymath} }
 \hfill
 \parbox{2cm}
  { \begin{equation}
     \label{phin}
    \end{equation}
  } \\
From (Eq.~\ref{Vcal})  the visibility
phase $\psi_{ik}$ on the baseline $i k$ will be related to the
observed phase $\varphi_{ik}$ by
 \begin{equation}
   \label{obsp}
    \varphi_{ik} = \psi_{ik} + \theta_i - \theta_k + \epsilon_{ik} \,,
 \end{equation}
where $\epsilon_{ik}$ is the phase noise.
Then the {\em closure\index{Phase!closure} phase} $\Psi_{ikl}$ around a closed
triangle of baseline $i k, k l, l i$,
 \begin{equation}
   \label{cltri}
    \Psi_{ikl} = \varphi_{ik} + \varphi_{kl} + \varphi_{li} =
                 \psi_{ik} + \psi_{kl} + \psi_{li} +
                 \epsilon_{ik} + \epsilon_{kl} + \epsilon_{li} \,,
 \end{equation}
will be independent of the phase shifts $\theta$ introduced by the
individual antennas and the time variations. With this procedure,  phase errors can be minimized. 
 
If four or more antennas are used simultaneously, then the 
{\em closure amplitudes} can be formed.  These are
independent of the antenna gain factors:
 \begin{equation}
   \label{rat-cl}
     A_{klmn} = \frac{|V_{kl}| |V_{mn}|}{|V_{km}| |V_{ln}|}  \; .
 \end{equation}
Both phase and closure amplitudes can be used to improve the quality of the
complex visibility function\index{Closure amplitude}.

  At each antenna there is  an unknown complex gain factor $g$ with
amplitude and phase, the total number of unknowns  can be
reduced significantly by measuring closure phases and amplitudes. If four
antennas are available, 50\,\% of the phase information and 33\,\% of
the amplitude information can thus be recovered; in a 10 antenna
configuration, these ratios are 80\,\% and 78\,\%
respectively.

 
\subsubsection{Calibrations, Gridding, FFTs, Weighting and Self Calibration }

For two antenna interferometers,  phase calibration can only be made pair-wise. This is referred to as $''$baseline based$''$ solutions for the calibration. 
For a multi-antenna system, 
 $''$antenna based$''$ solutions   are preferred.
These are determined by applying phase and amplitude closure for subsets of antennas and then solving for the best fit for each.


Normally the
Cooley\index{Cooley-Tukey Fast Fourier Transform}-Tukey fast\index{FFT}
Fourier transform algorithm is used to  invert  (Eq.~\ref{i-str})
To apply the simplest version of the FFT, the visibilities must
be placed on a regular grid with  sizes that are powers of two of
the
sampling interval. Since the data seldom lie on such regular
grids, an interpolation
scheme must be used. 
From the gridded $(u, v)$ data,  an image with a resolution
corresponding to $\lambda/D$, where $D$ is the array size,  is obtained. However, this may still contain artifacts
caused by the observing procedure, especially  the 
limited coverage of the ($u, v$) plane. Therefore the
dynamic
range of such so-called {\em dirty\index{Dirty map}} maps is rather
small.
This can be improved by further  analysis.

If the calibrated visibility function $V (u, v)$ is known for the full
$(u,v)$ plane both in amplitude and in phase, this can be used to
determine the modified  (i.e., structure on angular scales finer than $\lambda/D$ are lost) intensity distribution $ I' (x, y)$
by performing the Fourier transformation (Eq.~\ref{i-str}).
However, in a realistic situation $V (u, v)$ is only sampled at discrete
points 
and in some regions of the $(u,v)$ plane, $V(u, v)$ is not measured at all.
 The visibilities can be weighted  by a grading function, $g$. For a discrete
number of visibilities,  a version of (Eq.~\ref{i-str}) involving a
summation, not an integral, is  used to obtain an image with the use of a discrete Fourier
transform (DFT):
 \begin{equation}
   \label{i-strstr}
    I_{\rm D} (x, y) = \sum_k g (u_k, v_k) V (u_k, v_k) \expo^{-2 \pi \im         ( u_k x + v_k y)} \, ,
 \end{equation}
where $g (u, v)$ is a weighting function referred to as the 
grading\index{Grading} or apodisation\index{Apodisation}.
$g(u,v)$  can be used to change the effective beam
shape and side lobe level. There are two widely used weighting
functions: uniform and natural. Uniform \index{Weighting!uniform} weighting
uses $g(u_k,v_k)=1$, while
natural  \index{Weighting!natural} weighting uses
$g(u_k, g_k) = 1 / N_{\rm s}(k)$, where $N_{\rm s}(k)$ is
the number of data points within a symmetric region of the $(u,v)$
plane.
Data which are naturally weighted result in lower angular resolution but give a
better signal-to-noise ratio than uniform weighting. But these are 
only extreme cases.  Intermediate weighting schemes 
are  referred to as {\em robust} weighting. 

Often the reconstructed image $I_{\rm D}$
may  not be a particularly good representation of $I'$, but
these are related by: 
 \begin{equation}
   \label{i-xy}
    I_{\rm D} (x, y) = P_{\rm D} (x, y) \otimes I' (x, y) \,,
 \end {equation}
where $I'(x, y)$ is
the best representation of the source intensity   modified by the primary beam shape; it 
 contains only
those spatial frequencies $(u_k,v_k)$ where the visibility function has
been measured. (see  (Eq.~\ref{i-str})). The expression for $  P_{\rm D}$ is:  
 \begin{equation}
   \label{pd}
    P_{\rm D}        = \sum_k g (u_k, v_k) \expo^{- 2 \pi \im
                       ( u_k x + v_k y)}
 \end{equation}
this is the response to a point source, or the {\em point spread function}
 PSF\index{Interferometer!point spread function} for the dirty\index{Beam!dirty} beam. Thus   $P_{\rm D}$
is a transfer function that distorts the image;  $P_{\rm D}$ is produced assuming an amplitude of unity and phase zero at each point sampled. This is the
response of the interferometer system to a point source. The sum in (Eq.~\ref{pd}) extends 
over the same positions $(u_k, v_k)$ as in (Eq.~\ref{i-strstr});  the
sidelobe structure of the beam depends on the distribution of these
points.


Amplitude and
phase errors scatter power across the image, giving the appearance of
enhanced noise. This problem can be alleviated to an impressive
extent by the method of {\em self-calibration}. This process can be applied
if there is a sufficiently intense compact feature in the field contained
within the primary beam of the interferometer system. If self-calibration can be applied, the 
positional information is usually  lost. 
Self-calibration can be restricted  to an improvement of phase alone or
to both phase and amplitude. Normally, self-calibration  is carried in the
$(u,v)$ plane.  If this method is used on objects with low signal-to-noise
ratios,  this may lead to  a  concentration of  random
noise into one part of the interferometer image (see Cornwell \& Fomalont 1989).
%
For measurements of weak spectral lines,  self-calibration is  carried out using  a
continuum source in the field. The corrections are then applied to the spectral
line data. In the case of intense lines, one of the frequency channels
containing the emission is used.

\subsubsection{More Elaborate Improvements of Visibility Functions: The CLEANing Procedure}
{CLEAN}ing is the most commonly used technique to improve single
radio interferometer images (H\"ogbom 1974).
 In addition to its inherent low
dynamic range, the dirty map often contains features such as negative intensity artifacts that cannot be 
real.
Another unsatisfactory aspect is that the 
solution
is quite often rather unstable, in that it can change drastically
when more visibility data are added.

The CLEAN method approximates the  intensity distribution that represents the best image of the source (subject to angular resolution, noise, etc.), 
$ I (x, y)$,  by the superposition of a finite number
of point sources with positive intensity $A_i$ placed at positions
$(x_i, y_i)$.
 The goal  of CLEAN\index{CLEAN} to determine the $A_i( x_i, y_i)$, such that
 \begin{equation}
  \label{i-clean}
I'' (x,y) = \sum_i A_i \,P_{\rm D} (x - x_i, y - y_i) + I_\epsilon (x,
y) \,
 \end{equation}
where $I''$ is the dirty map obtained from the inversion
of the visibility function and $P_{\rm D}$ is the dirty beam (Eq.~\ref{pd}).
$I_\epsilon (x, y)$ is the residual\index{Brightness distribution!residual}
brightness distribution after
decomposition. Approximation (Eq.~\ref{i-clean}) is considered 
successful if $I_\epsilon$ is of the order of the noise in the measured
intensities. This decomposition must be carried out 
 iteratively.

The CLEAN algorithm is most commonly applied  in the image plane. This is
an iterative method which functions in the following fashion: (1)  find
the peak intensity of the dirty image,
then subtract a fraction $\gamma$ (the so-called $''$loop gain$''$)  having  the shape of the dirty beam
from the image, and (2)  repeat this $n$ times.

 This \index{CLEAN!loop gain} {\em loop gain} has values 
$0 < \gamma <  1$ while $n$ is often taken to be 10$^4$. The goal is that 
the intensities of the residuals are comparable to the noise limit. Finally, 
 the resulting  model is convolved with a
{\em clean\index{Clean beam} beam}  of
Gaussian shape with a FWHP given by the angular resolution expected from $\lambda/D$ where $D$ is the maximum baseline length.
Whether this algorithm produces a realistic image depends on the quality of the data and other variables.

 \subsubsection{More Elaborate Improvements of Visibility Functions: The Maximum Entropy  Procedure}
 
The Maximum Entropy Deconvolution Method (MEM)
is commonly used to produce a single optimal image from a set of
separate but contiguous images (Gull \& Daniell 1978).
The problem of how to select the $''$best$''$ image from many possible images
which all agree with the measured visibilities is solved by
MEM\index{MEM}. Using MEM, \index{Maximum entropy method} those values of the
interpolated visibilities are selected, so that the resulting image
is consistent with all  previous relevant data. In addition, the MEM image has
maximum smoothness. This is obtained by
maximizing the {\em entropy} of the image. One  definition of
entropy is given by
\begin{equation}
 \label{entropy}
   {\cal H } = - \sum_i I_i \left[ \ln \bigg(\frac{I_i}{M_i}\bigg) -1
\right] \,,
\end{equation}
where $I_i$ is the deconvolved intensity and $M_i$ is a reference image
incorporating all $''$a priori$''$ knowledge. In the simplest case $M_i$ is the
empty field $M_i = {\rm const} > 0$, or perhaps a lower angular
resolution image.
 
Additional constraints might require that all measured visibilities should
be reproduced exactly, but in the presence of noise such constraints are
often incompatible with $I_i > 0$ everywhere. Therefore the MEM image is
usually constrained to fit the data such that
\begin{equation}
 \chi ^2 = \sum \frac{|V_i - V_i'|^2}{\sigma_i^2}
\end{equation}
has the expected value, where $V_i$ is the measured visibility, $V_i'$
is a visibility corresponding to the MEM image and $\sigma_i$ is the
error of the measurement.

\vspace{5mm}
\noindent
 Acknowledgement: K. Weiler made a thorough review of the text and H. Bond suggested a number of improvements.

%
%

%
%

\end{document}